\documentclass[letterpaper,11pt]{article}
\usepackage{amsmath}
\usepackage{amsthm,amssymb, mathtools}

\usepackage[utf8]{inputenc} 
\usepackage[T1]{fontenc}    
\usepackage{url}            
\usepackage{booktabs}       
\usepackage{amsfonts}       
\usepackage{nicefrac}       
\usepackage{microtype}      
\usepackage{caption}
\usepackage{subcaption,comment}
\usepackage{xfrac}
\usepackage{fullpage}

\usepackage{amsmath}
\usepackage{amssymb, mathtools}
\usepackage{amsthm}
\usepackage{graphicx}
\usepackage{float}
\usepackage{amsfonts}
\usepackage{titlesec}
\usepackage[pdfstartview=FitH,colorlinks,linkcolor=blue,filecolor=blue,citecolor=blue,urlcolor=blue,pagebackref=true]{hyperref}

\usepackage{xcolor}
\usepackage{mathtools}

\newcommand{\Advan}{\mathrm{Adv}}

\newcommand{\sanjam}[1]{}
\newcommand{\sedits}[1]{}
\newcommand{\FT}[1]{}
\newcommand{\SJ}[1]{}

\newcommand{\Mnote}[1]{}
\newcommand{\MohNote}[1]{}

\newcommand{\Snote}[1]{}

\newcommand{\AbhrNote}[1]{}
\newcommand{\SamNote}[1]{}
\newcommand{\SSNote}[1]{}

\newcommand{\fhat}[2]{\ifthenelse{\equal{#2}{}}{\hat{f}(#1)}{\ifthenelse{\equal{#2}{0}}{\hat{f}(\emptyset)}{\hat{f}(#1_{\leq #2})}}}
\newcommand{\ftild}[2]{\ifthenelse{\equal{#2}{}}{\tilde{f}(#1)}{\ifthenelse{\equal{#2}{0}}{\tilde{f}(\emptyset)}{\tilde{f}(#1_{\leq #2})}}}
\newcommand{\ftildstar}[2]{\ifthenelse{\equal{#2}{}}{\tilde{f^*}(#1)}{\ifthenelse{\equal{#2}{0}}{\tilde{f^*}(\emptyset)}{\tilde{f^*}(#1_{\leq #2})}}}
\newcommand{\ghat}[2]{\ifthenelse{\equal{#2}{}}{\hat{g}(#1)}{\ifthenelse{\equal{#2}{0}}{\hat{g}(\emptyset)}{\hat{g}(#1_{\leq #2})}}}
\newcommand{\mfix}[2]{\ifthenelse{\equal{#2}{}}{m(#1)}{\ifthenelse{\equal{#2}{0}}{m(\emptyset)}{m(#1_{\leq #2})}}}

\newcommand{\fapp}[2]{\ifthenelse{\equal{#2}{}}{\tilde{f}(#1)}{\ifthenelse{\equal{#2}{0}}{\tilde{f}(\emptyset)}{\tilde{f}(#1_{\leq #2})}}}

\newcommand{\paragcompact}[1]{\noindent{\bf #1}}

\newcommand{\Risk}{\mathsf{Risk}}

\newcommand{\aSF}{\mathsf{a}}
\newcommand{\gSF}{\mathsf{g}}
\newcommand{\hSF}{\mathsf{h}}

\newcommand{\avr}[2]{\ifthenelse{\equal{#2}{}}{\aSF({#1})}{\ifthenelse{\equal{#2}{0}}{\aSF(\emptyset)}{\aSF({#1}_{\leq #2})}}}

\newcommand{\avrMax}[2]{\ifthenelse{\equal{#2}{}}{\aSF^*({#1})}{\ifthenelse{\equal{#2}{0}}{\aSF^*(\emptyset)}{\aSF^*({#1}_{\leq #2})}}}

\newcommand{\avrApp}[2]{\ifthenelse{\equal{#2}{}}{\tilde{\aSF}({#1})}{\ifthenelse{\equal{#2}{0}}{\tilde{\aSF}(\emptyset)}{\tilde{\aSF}({#1}_{\leq #2})}}}

\newcommand{\avrAppMax}[2]{\ifthenelse{\equal{#2}{}}{\tilde{\aSF}^*({#1})}{\ifthenelse{\equal{#2}{0}}{\tilde{\aSF}^*(\emptyset)}{\tilde{\aSF}^*({#1}_{\leq #2})}}}

\newcommand{\ArgMax}[2]{\ifthenelse{\equal{#2}{}}{\hSF({#1})}{\ifthenelse{\equal{#2}{0}}{\hSF(\emptyset)}{\hSF({#1}_{\leq #2})}}}

\newcommand{\AppArgMax}[2]{\ifthenelse{\equal{#2}{}}{\tilde{\hSF}({#1})}{\ifthenelse{\equal{#2}{0}}{\tilde{\hSF}(\emptyset)}{\tilde{\hSF}({#1}_{\leq #2})}}}

\newcommand{\gain}[2]{\ifthenelse{\equal{#2}{}}{\gSF(#1)}{\gSF(#1_{\leq #2})}}
\newcommand{\gainMax}[2]{\ifthenelse{\equal{#2}{}}{\gSF^*(#1)}{\gSF^*(#1_{\leq #2})}}
\newcommand{\gainApp}[2]{\ifthenelse{\equal{#2}{}}{\tilde{\gSF}(#1)}{\tilde{\gSF}(#1_{\leq #2})}}
\newcommand{\gainAppMax}[2]{\ifthenelse{\equal{#2}{}}{\tilde{\gSF}^*(#1)}{\tilde{\gSF}^*(#1_{\leq #2})}}

\newcommand{\pr}[2][]{\Pr_{\ifthenelse{\isempty{#1}}{}{{#1}}}\left[{#2}\right]}

\newcommand{\sm}{\setminus}

\newcommand{\remove}[1]{}

\newcommand{\set}[1]{\left\{ #1 \right\}}

\newcommand{\bits}{\{0,1\}}

\newcommand{\R}{{\mathbb R}}
\newcommand{\N}{{\mathbb N}}

\newcommand{\cD}{{\mathcal D}}

\newcommand{\cX}{{\mathcal X}}

\newcommand{\bfx}{\mathbf{x}}

\newcommand{\eps}{\varepsilon}

\newcommand{\Exp}{\operatorname*{\mathbb{E}}}
\newcommand{\Ex}{\Exp}

\newcommand{\maj}{\operatorname*{maj}}

\newtheorem{theorem}{Theorem}

\newtheorem{definition}{Definition}

\newcommand{\sdotfill}{\textcolor[rgb]{0.8,0.8,0.8}{\dotfill}} 

\makeatletter
\def\th@protocol{%
    \normalfont 
    \setbeamercolor{block title example}{bg=orange,fg=white}
    \setbeamercolor{block body example}{bg=orange!20,fg=black}
    \def\inserttheoremblockenv{exampleblock}
  }
\makeatother

\newtheorem{proto}[theorem]{Protocol}
\newtheorem{protoc}[theorem]{Protocol}

\newcommand{\namedref}[2]{#1~\ref{#2}}

\newcommand{\torestate}[3]{%
\expandafter \def \csname BBRESTATE #2 \endcsname{#3}
\theoremstyle{plain}
\newtheorem{BBRESTATETHMNUM#2}[theorem]{#1}
\begin{BBRESTATETHMNUM#2}\label{#2}\csname BBRESTATE #2 \endcsname   \end{BBRESTATETHMNUM#2}
\newtheorem*{BBRESTATETHMNONNUM#2}{\namedref{#1}{#2}}
}

\newcommand{\restate}[1]{\begin{BBRESTATETHMNONNUM#1}[Restated] \csname BBRESTATE #1 \endcsname
\end{BBRESTATETHMNONNUM#1}}

\begin{document}
\title{Is Private Learning Possible with Instance Encoding?}
\author{
Nicholas Carlini\\
\texttt{ncarlini@google.com}
\and
Samuel Deng\\   
\texttt{sd3013@columbia.edu} 
\and
Sanjam Garg\\
\texttt{sanjamg@berkeley.edu} \\
\and
Somesh Jha\\
\texttt{	jha@cs.wisc.edu} \\
\and
Saeed Mahloujifar\\
\texttt{sfar@princeton.edu}\\
\and
Mohammad Mahmoody\\
\texttt{mohammad@virginia.edu}\\
\and
Shuang Song\\
\texttt{shuangsong@google.com }\\
\and
Abhradeep Thakurta\\
\texttt{athakurta@google.com}\\
\and
Florian Tram\`er\\
\texttt{tramer@cs.stanford.edu}\\
}
\date{}
\maketitle
\setlength{\tabcolsep}{11pt}

\begin{abstract}
A private machine learning algorithm hides as much as possible about its training data while still preserving accuracy. In this work, we study whether a non-private learning algorithm can be made private by relying on an instance-encoding mechanism that modifies the training inputs before feeding them to a normal learner. We formalize both the notion of instance encoding and its privacy by providing two attack models. We first prove impossibility results for achieving a (stronger) model. Next, we demonstrate practical attacks in the second (weaker) attack model on InstaHide, a recent proposal by Huang, Song, Li and Arora [ICML'20] that aims to use instance encoding for privacy.
\end{abstract}

\newcommand\blfootnote[1]{%
  \begingroup
  \renewcommand\thefootnote{}\footnote{#1}%
  \addtocounter{footnote}{-1}%
  \endgroup
}
\blfootnote{$^*$ Authors ordered alphabetically.} 

\section{Introduction}

Neural networks are increasingly trained on sensitive user data, 
for example building classifiers to diagnose diseases from medical images~\cite{hosny2018artificial,wernick2010machine}
or help users compose emails or text messages by training on actual user data~\cite{dai2019gmail}.

Protecting the privacy of users' data while training such models currently requires either a trusted central party with all users' data, or applying cryptographic techniques such as multiparty computation~\cite{mohassel2017secureml, mohassel2018aby3, bonawitz2016practical} 
that introduce large computation and communication overheads.
In turn, preventing the trained model itself from leaking private information, e.g., with differential privacy~\cite{dwork2006calibrating,chaudhuri2011differentially,abadi2016deep}, typically comes at a high cost in accuracy.
This raises the question: 
\emph{Are there other ways to perform private learning without sacrificing performance or accuracy?}

An alternate method for privately training
a neural network is to first convert users' data to an encoded (private) version, 
and then train a non-private model on this encoded dataset~\cite{huang2020instahide,raynal2020image}.
Since the training data has been privately encoded, the model training gets privacy ``for free.''
We formalize this \emph{private instance encoding} setup, and investigate fundamental limits on how well such an approach can work in theory.

We show that training a model on encoded data cannot offer privacy guarantees as strong as cryptographic techniques. Specifically, we prove  that no useful encoding can resist \textbf{distinguishing} attacks of two forms. Our first attack distinguishes with non-negligible probability whether dataset $S_1$ or dataset $S_2$ was used to generate an encoded dataset. Our second attack distinguishes encodings of \emph{instances} alone with a higher probability by relying on further assumptions about the encoding function and its utility. We formalize these definitions in Section~\ref{sec:defs}, and theorems in Section~\ref{sec:impossibilities}.

We next study practical private instance encoding schemes. While our distinguishing attacks apply to any instantiation of instance encoding, we now attempt the stronger goal of \textbf{reconstruction} for specific instance encoding schemes.
Given the encoded dataset, a reconstruction attack recovers (nearly identical) copies of individual training examples used.
This privacy goal is weaker than indistinguishability, and arguably the weakest form of privacy that could be expected.

We design a reconstruction attack that breaks InstaHide~\cite{huang2020instahide}, 
the state-of-the-art privacy-preserving encoding-based technique
which was awarded a Bell Labs Prize~\cite{disgrace2020}.
InstaHide applies to image classification. Its encoding function \emph{mixes} together   multiple images~\cite{zhang2017mixup} (with a linear pixel blend), and then it randomly flips the signs of the pixels.  
Our reconstruction attack (Section~\ref{sec:instahide_attack}) recovers high-quality reconstructions---for example we solve the challenge released by the authors \cite{challenge} and recover a nearly visually identical reconstruction of all the private encoded images. Our attack demonstrates that InstaHide fails to satisfy meaningful privacy notions.

Our attack leverages the fact that InstaHide encodings \emph{are} distinguishable, as our theoretical results predict.
Given multiple encoded images (produced by training for multiple epochs), we cluster encodings that correspond to the same source image. We then merge these encodings to recover the original image by solving a noisy linear system.
Our attack sidesteps the encoding's sign flipping (which provides no privacy in itself) by simply taking the absolute value before all operations.

We further show (Section~\ref{sec:instahide_parameters}) that  extensions of InstaHide offer no more privacy.
Mixing \emph{more} images into each encoding \emph{strengthens} our attack. Even given a \emph{single} encoding of an image, an attacker with precise knowledge of InstaHide's \emph{public} parameters can reconstruct the encoded image near-perfectly.


\subsection{The Instance Encoding Problem}
\label{ssec:instance_encoding}

In the \emph{instance encoding} problem setup, the defender encodes a (sensitive) training dataset $S=\set{(x_1,y_1),\dots(x_n,y_n)}$ by processing it with an encoding function $E$.
The encoded version $ \Tilde{S} \gets E(S)$ is released publicly.

Any learning algorithm $L$ can then train on the encoded set $\tilde{S}$ to learn the \emph{concept function} $c$ that was used to construct the training dataset 
(i.e., it was used to construct the labels $c(x_i) \equiv y_i$).
The encoding is useful if get both of the following.
\begin{itemize}
    \item \textbf{Utility-preserving.} A model trained on the encoded datset $\Tilde{S}$ should
    be (approximately) as accurate as a model trained on the original dataset $S$.
    \item \textbf{Privacy-preserving.} Given access to the encoded dataset $\Tilde{S}$, it should
    be difficult to learn sensitive properties about the original training dataset $S$.
    
\end{itemize}


\paragraph{Forms of encoding.}
An encoding function is an arbitrary function operating over a training dataset $S$, allowing
for a wide range of techniques.
At one extreme, a valid encoding function could take the entire training dataset $S$, 
run the learning algorithm $L$ on all of it, and  output the trained model $h \gets L(S)$ as the output
of the encoding.
This ``encoding'' may (or may not) be useful or private.
At the other extreme, an encoding scheme might be completely ``local'' and operate \emph{independently} 
on each training example to produce $\Tilde{S} = \{e(x_i, y_i)\,\, \colon (x_i,y_i) \in S\}$.

Another natural class of encoding schemes are those that are based on ``mix-up''-type operations~\cite{zhang2017mixup} that apply a simple linear operation on a small number of instances to produce an encoded instance. Such encoding schemes usually have a nice property: they can be ``decomposed'' into two encoding algorithms that operate separately on instances and on labels.  This class of encoding schemes includes the mix-up encoding function used in the recent InstaHide protocol \cite{huang2020instahide}. Since decomposable encodings apply to instances and labels separately, in such cases we indeed deal with an \emph{instance} encoding together with a \emph{label} encoding that work in tandem.

\paragraph{Adversary capabilities.}
We assume the adversary is given access to the encoded dataset $\Tilde{S}$.
The adversary does not have \emph{any} access to the original dataset $S$
In some algorithms, the encoding scheme might receive as input some public data $P$; for these schemes we assume the adversary has access to $P$. (The InstaHide algorithm, for example, takes a ``private'' and ``public'' dataset as input.)
 
\paragraph{Adversary objective.}
The adversary aims to learn as much information as possible about $S$ given all available information.
The most powerful attacks we consider are complete \emph{reconstruction} attacks that recover training examples $x'_i$ where $x'_i \approx_m x_i$ according to some similarity metric $m$ (e.g., Euclidean distance).

We also consider more restrictive \emph{distinguishing} (inference) attacks where the adversary aims only to determine if a particular $x_i$ was used as training data or not.

\vspace{1em}
\paragraph{Main question.} 
This paper studies the following question
\begingroup
\addtolength\leftmargini{-0.1in}
\begin{quote}
    \emph{For a dataset $S$ with instances labeled by a nontrivial concept function,
    is it possible to design an encoding function $E(S)=\Tilde{S}$ so that given $\Tilde{S}$
    a learning algorithm $L$ can produce an accurate model
    but so that the original data $S$ remains hidden from an adversary?}
\end{quote}
\endgroup
Note that we assume that the adversary has direct access to $\tilde{S} \gets E(S)$, before the learning algorithm $L$ is run on $\tilde{S}$. 

It is easy to achieve privacy alone if $E$ hides everything about $S$ (e.g., define $E(x) \equiv 0$ as a constant function), but then no meaningful learning is possible.
Alternatively, if the concept function is trivial (e.g., all examples
have the same label) then trivial encoding functions exist.
We are interested only in encoding functions that operate over nontrivial concept functions.
Our goal is to understand the barriers and trade-offs that arise between the privacy provided by the encoding function vs. the \emph{utility/accuracy} of the learning algorithm.

\subsection{Results}
\label{ssec:overview}
We provide negative results in the form of theoretical barriers that prevent
any encoding function from protecting some forms of privacy. Moreover, we demonstrate practical
attacks on specific encoding functions from the literature \cite{huang2020instahide}.

\subsubsection{Theoretical Impossibility Results.}
We prove that for the case of distinguishing attacks, it is not possible to
construct nontrivial encoding functions that preserve both utility and
a weak form of distinguishing privacy.

\paragraph{Limits of privacy with dataset encoding.} We first study \emph{distinguishing} attacks whose goal is to find out with probability (non-negligibly) more than $\sfrac12$ which dataset out of two (different) sets $S_1,S_2$ has been   encoded. In fact, we will show how to achieve this even if the datasets share many similarities and only differ in one example pair (with the \emph{same} label).
\begin{theorem}[Informally stated -- limits of privacy based on dataset encoding] \label{thm:Inf1} Let $(E,L)$ be an arbitrary encoding and learning scheme with  at least $51\%$ accuracy on the original data (not the encoded data).
Then for any ``nontrivial'' concept class $C$ (e.g., sufficient to be closed under complement and to contain at least two distinct concepts), there is an adversary who can pick a concept $c \gets C$ and two  datasets $S_1,S_2$  and distinguish their encodings with probability $1/2 + \Omega(1/n)$, while the sets satisfy the following restrictions:
\begin{enumerate}
    \item $S_1=\set{e_1}\cup S,S_2=\set{e_2}\cup S$ differ in one sample only.
    \item All instances in $S_1,S_2$ are cleanly labeled (by  $c$). This includes also the  differing examples $e_1,e_2$ (i.e., $e_1=(x_1,y),e_2=(x_2,y)$ for $y=c(x_1)=c(x_2)$).
\end{enumerate} 
\paragraph{Limits of privacy with instance encoding.} We now describe our next result which states the limits of what \emph{instance} encoding (as a special form of general dataset encoding) can offer for (our minimal and natural indistinguishability-based notion of) data privacy. In this setting, we deal with a \emph{decomposable} encoding, which encodes instances and their labels separately. Such decomposable encodings can cover, e.g., the mix-up operation \cite{zhang2017mixup} used in InstaHide \cite{huang2020instahide}. For encoding $\tilde{x}$ and instance $x$, we write $x \in E^{-1}(\tilde{x})$ if $x$ is one of the instances that are used for generating the encoding $\tilde{x}$.
\end{theorem}
In our next result we show barriers for achieving privacy based on instance encoding, when two conditions hold: (1) The goal of the adversary is to distinguish encodings $\tilde{x}$ where $x \in E^{-1}(\tilde{x})$ from those where $x' \in E^{-1}(\tilde{x})$ for $x \neq x'$. (2) The learning algorithm $(E,L)$ allows some nontrivial accuracy on \emph{encoded} strings as defined above.

\begin{theorem}[Informally stated -- limits of privacy based on instance encoding] \label{thm:Inf2}
Suppose the goal is to learn instances that are distributed according to distribution $\cD$ and  the concept class is rich enough to contain $m$ concept functions $c_1,\dots,c_m$ that are each balanced under $\cD$ (i.e., $\Pr[c_i(\cD)=1]=1/2$) and are also independent from each other. (For example, this would be the case when the concepts contain $m$ orthogonal half spaces and $\cD$ is the isotropic Gaussian, all in  dimension $m$). Also, suppose the protocol $(E,L)$ has \emph{encoded} accuracy $1/2 + \delta$ for a constant $\delta > 0$. Then, the adversary can distinguish the encodings of two randomly selected instances $x,x' \gets D$ with probability  at least $0.99-O(\frac{1-2\delta}{1-\nicefrac{2}{\sqrt{m}}})$ (over the trivial bound of $1/2$).
\end{theorem}
We also prove a variant of Theorem \ref{thm:Inf2} that does  \emph{not} rely on the richness of the concept class. This result states that if instance encoding works on a single concept function $c$, then one of the following happens: \emph{either} (1) we obtain a distinguishing attack on the instance encoding, or (2) the learning error on $c$ can be arbitrarily close to $0$. This barrier applies to any setting where classifiers on $c$ achieve accuracy bounded away (by some constant) from $1$ (e.g., image classification).

\subsubsection{Concrete Attack Results.}

\begin{figure}[t]
    \centering
    \includegraphics[width=\columnwidth]{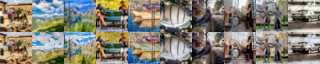}
    \caption{Our reconstruction attack on the InstaHide Challenge,
    for 10 randomly selected images \cite{challenge}.
    {\it Upper row:} ground truth obtained from a cryptanalytic attack \cite{kelsey1998cryptanalytic} on the PRNG in InstaHide's implementation (Appendix~\ref{apx:prng_attack}). 
    {\it Lower row:} our reconstruction attack yields high fidelity image reconstructions.
    A complete set of the 100 recovered images are in Appendix~\ref{apx:figures}.} 
    \label{fig:instahide_break}
\end{figure}

We further demonstrate that InstaHide \cite{huang2020instahide}, a practical instance encoding scheme,
is not private.
Figure~\ref{fig:instahide_break} shows the result of our attack on the InstaHide Challenge.
This challenge contains $\lvert\Tilde{S}\rvert = 5,000$ encoded images
from $\lvert S \rvert = 100$ original encoded images---thus, each original image
has been encoded 50 different times.
We are able to completely reconstruct a nearly-identical version $S$ given access to $\Tilde{S}$.

Our attack directly leverages the fact that InstaHide encodings \emph{are}
distinguishable.
Given the encoded dataset, we construct a similarity function that allows us to
detect when two examples $x,y \in \Tilde{S}$ are derived from the same original image in $S$. Theorem \ref{thm:Inf2} explains why such similarity function should exist as we can use a rich concept class to map encoding to a embedding space and use clustering to identify encoding that encode the same image. However our actual attack takes a different approach and leaves the computation of this similarity metric to a neural network. Specifically, we train a neural network that distinguishes whether a pair of encodings share the same input image which generalizes to unseen examples with high accuracy. 
This construction already consists of a privacy leak according to the definition in the prior section.
However, we are able to extend the attack to complete reconstruction.
Given our similarity function, we can group together multiple encoded images $T \subset \Tilde{S}$
so that all images in the encoded subset $T$ correspond to the same original image.
and then develop a recovery function $r$ so that $r(T) \approx x \in S$.

We further introduce a second attack that works in \emph{linear} time
and that succeeds even when given a \emph{single} encoding of an image $x_i$. This attack assumes knowledge of the public images used in
the InstaHide algorithm. (While we assume an adversary would have access to the ``public'' images, we can not use this attack on the InstaHide challenge as it \emph{does not} release these images.)
This attack similarly produces nearly perfect image reconstructions when it succeeds, but does fail with a small constant probability in our experiments.

\subsection{Related Work}
Theorem~\ref{thm:Inf1}  can be seen as a  (dimension-independent) lower bound on the sample complexity of private PAC learning.  Prior work has studied similar lower bounds on sample complexity of learning algorithms in various contexts. For example, the work of~\cite{duchi2014local,hardt2010geometry, bun2016order} use packing arguments to give sharp bounds for  parameter/probability
estimation goals, and~\cite{bun2018fingerprinting} proves lower bounds on the sample complexity of differentially private algorithms that accurately answer large sets of counting queries. In addition, lower bounds on the sample complexity of  differentially private (general) PAC learning were proved in \cite{bassily2014private,beimel2014bounds}. It might be possible to improve Theorem~\ref{thm:Inf1} by incorporating the data dimension, however not depending on the dimension is a postive, and we emphasize that our result comes with specific guarantees that are important: the two sets are consistent with a concept function and have the same set of labels. This makes our lower bound more amenable to real world setting, where we want to distinguish two data sets with say, the same number of cat and dog images in them.

Our attacks on data privacy of ML models are related to ``membership inference'' attacks~\cite{shokri2017membership,long2017towards,salem2019ml,long2018understanding} as well as  \emph{model inversion} attacks \cite{fredrikson2014privacy,fredrikson2015model,wu2016methodology}, and our attack on InstaHide is a form of reconstruction attack
\cite{dinur2003revealing,dwork2017exposed,backes2016membership,dwork2015robust,sankararaman2009genomic,homer2008resolving}.

\section{Privacy with Instance Encoding: Definitions \Snote{Should we rename this section to "Privacy with Instance Encoding: Threat model" and then have "privacy definitions" as a subsection. }}
\label{sec:threat}

\label{sec:defs}
\subsection{Formal Definitions For Learning with Instance Encoding}
\paragcompact{Notation.} Let $X$ be an instance space and $Y$ be a label space. We specify a learning problem with a  tuple $(\cD, C, H)$ where $C \subset Y^X$ (resp. $H \subset Y^X$) is a class of concept (resp. hypothesis) functions from $X$ to $Y$  and $\cD$ is a   distributions over $X$.\footnote{Since we aim to prove impossibility results,    focusing on the \emph{distribution-specific} learning setting  only makes our results stronger.}
For a concept function $c\in C$, we use $\cD_c$ to specify the joint distribution of labels and instances $(x, c(x))_{x \gets \cD}$ where we sample $x \gets \cD$ first, and then label $x$ according to $c$.   For a hypothesis $h \in H$, a concept class $c \in C$ and  $h$ with respect to $c$ under the distribution $\cD$ is defined as $\Risk(h,c)=\Pr_{x \gets \cD}[h(x)\neq c(x)]$. 

The following definition formalizes a general notion of encoding that allows instance encodings to depend on the   dataset. This, e.g., can capture encoding through data augmentation.
\begin{definition}[Dataset encoding mechanism] \label{def:enc}
A dataset encoding mechanism for a learning problem $(\cD, C,H)$ is a potentially randomized algorithm $E\colon (X\times Y)^* \to (\tilde{X}\times \tilde{Y})^*$ that takes a dataset $S$ as input and outputs an encoded dataset $\tilde{S}$. We define two properties for such encodings:

\begin{enumerate}
    \item {Decomposablity:} The encoding is decomposable if it performs on instances and labels separately; namely, it could be expressed using a pair of potentially randomized algorithms $E_X\colon X^* \to \tilde{X}^*$ and $E_Y\colon Y^* \to \tilde{Y}^*$ that share randomness. To encode a labeled dataset using such mechanism, one would apply $E_X$ to instances to get $\tilde{x}_1,\dots,\tilde{x}_m$ and $E_Y$ to labels to get $\tilde{y}_1,\dots,\tilde{y}_m$ and then output $\set{(\tilde{x}_1,\tilde{y}_1),\dots,(\tilde{x}_n,\tilde{y}_n)}$. Since such dataset encoding mechanism works on instances and labels separately, we refer to it as \emph{instance encoding} as well.
    \item {Locality:} And encoding scheme is $r$-local if all $ \tilde{x}\in E(S)$ would depend only on the randomness of $E$ and at most $r$ examples in $S$. If $\tilde{z}$ is an encoding that might depend on example $z$, we denote it by $z \in E^{-1}(\tilde{z})$.   Additionally, for $i \in [m]$, by $E^i(z_1,\dots,z_n)$ we denote the process of encoding $S$ using $E$ and then outputting one of the encoded examples $ \tilde{z} $ where $z_i\in E^{-1}(z)$   uniformly at random. 
    For decomposable encodings, we   define  notations   $E_X^i, E_Y^i$ for $i \in [m] \cup \set {-1}$ similarly. 
\end{enumerate}

\end{definition}


\paragraph{Examples.} We recall three natural  examples: (i) \emph{Identity mechanism:} In this case, we let $E$ be the identity function. This trivial encoding mechanism fully preserves the utility of learning on the original data set, but it  does not offer any privacy gains. (ii) \emph{Null mechanism:} Here, we let $E$ be   the constant $\bot$ function. In this case, the encoding  hides everything about the original data, but the generated encodings are useless for nontrivial training. (iii) \emph{Local DP mechanism:} Here, $E(S)$ generates a deferentially private noisy version of $S$. In this case, we can train using the encoded dataset with some possible degradation in accuracy. Note that in all these examples, encodings can be made decomposable and $1$-local 


\paragraph{Discussion.}
Definition \ref{def:enc} captures a broad range of techniques to achieve privacy. For example, it captures local (by choosing $r=1$) and central (by choosing $r=n$) encodings that might offer respectively local or central notions of differential  (or another form of) privacy.\footnote{A more general notion of locality refers to the setting where the data $S$ is partitioned into $r$ subsets, and then each of these subsets are independently encoded. Our $1$-locality definition covers this case when each of the sets includes one example only, but the definition could be generalized easily.}
Importantly, this encoding mechanism also captures InstaHide as it is allowed to be randomized and we put no limitation on the complexity of the encoding mechanism. Indeed, the InstaHide scheme is allowed to use randomness and also have access to a public dataset.
To incorporate InstaHide into our setting, the encoding algorithm could have the full public dataset hard-coded in its description and then use randomness to sample points from that dataset. In fact, InstaHide comes with decomposaibility and locality properties and is a special case of our definition. 


 We now formalize several accuracy and privacy notions of   encoding-based learning protocols. One can define accuracy on both encoded and original examples. Here we first define the accuracy on the original examples. 

\begin{definition}[Accuracy on plain (non-encoded) data] \label{def:accuracyPlain}
    The protocol $(E,L)$ is $(\eps,\delta)$-accurate, if for all $c\in C, n\in \N$,
\begin{align*}
    \Pr_{\substack{S\gets \cD_c^n, \tilde{S}\gets E(S), h\gets L(\tilde{S})}}[\Risk_{\cD}(h,c)\geq \eps(n)]\leq \delta(n).
\end{align*}
\end{definition}
%
%
%


One natural property that an instance encoding mechanism can provide is to enable the trained model to have some (perhaps weak) form of accuracy for predicting labels on the  \emph{encoded examples}. For example, suppose we use an $r$-local instance encoding mechanism, and that $\tilde{x}$ is an encoded instance that depends on $r$ distinct training samples, one of which is $x$. Then we could ask the trained model $h$ to  predict the \emph{true} concept $c(x)$ of $x$ when it is given the encoded sample $\tilde{x}$ as input. 
Indeed, we define (see Definition~\ref{def:accuracyEnc}) the  notion of \emph{encoded accuracy} for the model  $h$ to be  the probability of satisfying $h(\tilde{x})=c(x))$ when $x$ is a random instance and $x \in E^{-1}_X(\tilde{x})$. Of course this notion of accuracy may only be satisfiable in a \emph{weak} sense, as each locally encoded instance encoding $\tilde{x}$ depends on $r$ different instances which  may have different labels.
However, we argue that natural instance encoding schemes could still allow the error to be bounded away from (and smaller than) $1/2$. For example, using a $2$-local encoding on all pairs $(x,x')$ of instances in a set $S$ potentially allows  getting (weak) accuracy on encoded instances of $\approx 0.75$, because when the labels of $(x,x')$ are the same, the prediction of the model $h$ on the encoded string $\tilde{x}$ could be close to $1$, and in other cases it could be close to $0.5$. 

\begin{definition}[Accuracy on  encoded instances] \label{def:accuracyEnc}  We say the protocol $(E,L)$ is $(\eps, \delta)$-accurate on encoded instances if:
\begin{align*}
    \Pr_{\substack{S\gets \cD_c^n\\ \tilde{S}\gets E(S)\\ h\gets L(\tilde{S})}}\Bigg[\Pr_{\substack{\bfx \gets \cD^{n-1}\\ x\gets \cD\\ \tilde{x}\gets E^{1}_X(x,\bfx)}}[h(\tilde{x})\neq c(x)]\geq \eps(n)\Bigg]\leq \delta(n).
\end{align*}
Additionally we say the protocol has \emph{balanced} $(\eps,\delta)$-accuracy if for all possible labels $y$ we have
\begin{align*}
    \Pr_{\substack{S\gets \cD_c^n\\ \tilde{S}\gets E(S)\\ h\gets L(\tilde{S})}}\Bigg[\Pr_{\substack{\bfx \gets \cD^{n-1}\\ x\gets \cD\mid c(\cD)=y\\ \tilde{x}\gets E^{1}_X(x,\bfx)}}[h(\tilde{x})\neq c(x)]\geq \eps(n)\Bigg]\leq \delta(n).
\end{align*}
\end{definition}
Note that if the  decomposable encoding $E$ combines inputs with different labels, we might not expect the labeling error $\eps$ on encoded instances to be too close to $0$. Indeed, if an encoded instance $\tilde{x}$ combines two samples of different labels, the learned model necessarily assigns an ``incorrect'' label with respect to one of the instances. Nevertheless, if the encoder samples the $r$ inputs to combine uniformly at random, these $r$ inputs will have consistent labels with probability $2^{-r+1}$ and thus non-trivial accuracy is  possible whenever $r$ is constant. 

\subsection{Threat Model Formalization}
We now formalize our threat model introduced in Section~\ref{ssec:instance_encoding}.

\paragraph{Attacking in  polynomial time.} There is an asymmetry between the ``efficiency'' requirements for algorithms that are used frequently by \emph{hon} parties in a system, versus for algorithms that might rarely be used by malicious parties. When designing a learning scheme, one goal is to minimize its running time as much as possible. Even shaving a logarithmic factor might be important when the algorithm is run frequently and on large inputs. Attacks, on the other hand, are run rarely and in extreme cases (possibly only once). Thus, the system designer's goal is to achieve security against adversaries who might spend an \emph{unspecified}, yet  feasible, amount of resources. The reason is that we do not want to base its security on the hope that an adversary's running time cannot be improved further in the future. Indeed, modeling adversaries a polynomial-time entities  is commonplace in cryptography. Here we employ the same approach for   adversaries and the threat threat. Hence,   we  consider an attack efficient if it runs in polynomial time. Yet, we emphasize that  our attacks do have \emph{small} (absolute) running times, even though we do not optimize them.

\paragraph{Distinguishing  vs. reconstruction attacks.} Just like in encryption, our ultimate goal in private learning is to hide examples from the parties who are not supposed to know them. In both contexts, one can imagine weaker forms of attackers who can only \emph{distinguish} the target piece of data (e.g., plaintext in cryptography or private data in the context of learning) from irrelevant (e.g., random) pieces of information. This types of attacks, e.g., are the standard attacks against \emph{pseudorandom generators} in cryptography as well as attacks on differential privacy (e.g., membership inference attacks) in learning. A stronger, and more devastating form of attack consists of adversaries who \emph{completely recover} the sensitive information. E.g., one-way functions are design with respect to such attackers (and not surprisingly inverting functions breaks their pseudo-randomness as well). Such attacks also exist in the context of learning and, more generally, releasing public information about private. In this work we use \emph{both} types of distinguishing and reconstruction attacks. We prove \emph{general} barriers against distinguishing adversaries in the context of private-learning using instance encodings, and for the concrete case of InstaHide scheme, we present the (stronger) form of  adversaries, namely a  reconstruction attack.

\paragraph{What does it mean to keep examples private?} 
In full generality, a multi-party learning protocol consists of a set of parties $P_1,\dots, P_n$. Each $P_i$ has access to a dataset $S_i$ that they use for training. We refer to the transcript of communication between the parties as $T$ and the output of the protocol as $M$. The parties can also have some secret randomness $R_1, \dots, R_n$. Within this setting, we can define two types of privacy that are both important and complementary.

\remove{
We also note that the dataset encoding framework differs from the MPC setting in terms of the protocol output. Specifically, in MPC protocols the output resulting from the computation is exact. On the other hand, dataset encoding framework allows for some inaccuracy in the output, i.e. the trained model, of the aggregator. While this situation differs a bit, the same intuitive argument for impossibility applies. Namely, by encoding one's input, others can still enumerate their inputs (beyond their honest inputs) and observe all possible outcomes that the joint computation could produce. 
}

 \emph{Physical privacy (MPC).}
    In this setting, there is a set of indices of honest parties $I_{hon}$ that act based on the rules of the protocol. There is a set of indices $I_{dh} = [n] \sm I_{hon}$ that indicates the set of parties that are dishonest. The privacy of the scheme requires that no polynomial-time adversarial algorithm $A$ who completely controls the parties in $I_{dh}$ cannot extract any information about $S_{I_{hon}}$  other that what one can infer by only looking at the output of the protocol (which is the final model in case of multi-party learning). 
    %
     
    %
    Note that in this setting, the privacy requirement does not capture leakage from the actual outcome of the protocol. For example, one can imagine a protocol that outputs the training data of all the parties, while still satisfying physical privacy trivially. Therefore, ultimately, physical privacy shall be accompanied also by a leakage analysis of the final output. 

\emph{Functional privacy.}
    Here, 
    again the goal of the adversary is to infer some sensitive information about $S_{I_{hon}}$, but mainly by looking at the at the output $M$. Note that here adversary's goal is not to gain some \emph{extra} knowledge about $S_{I_{hon}}$ beyond what $T$ entails, but rather to find out something about $S_{I_{hon}}$ based on $M$ compared to when $M$ is not known. 
    Indeed, notions such as differential privacy or  $k$-anonymity are invented to allow us quantify the functional form of privacy.
    To achieve functional privacy in contexts such as searchable encryption, sometimes a  leakage function $Leakage(M,R_{I_{dh}},S_{I_{dh}})$ is defined to model what is considered acceptable to be leaked to the adversary. 
    
    We emphasize that the above two types of privacy are incomparable and complementary.

\paragraph{Can instance encoding provide physical privacy?}
 Private learning with $1$-local instance encoding can be seen as a protocol where each party sends only one message non-interactively. Then, using these messages, the protocol outputs a model $M$. Now, one can try to prove both physical and functional privacy for such a protocol.





We first observe that no dataset encoding algorithm $E$ achieves the physical privacy required by an MPC protocol, unless the learning task is trivial (i.e., does not depend on the data) or the learning algorithm is run by a trusted party. This follows from a folklore claim (proven in \cite{HLP11}) that it is impossible to construct an MPC protocol where parties send only one message each---represented by the encoded dataset sent by each of the parties. We now give an intuition of this claim, tailored to the two-party case of our dataset encoding framework.
In the two-party case, computation proceeds as follows: Each party encodes its dataset $S_i$ to $E(S_i)$ and sends it to an aggregator. Next, the aggregator performs the training directly on the encoded datasets $E(S_1), E(S_2)$, yielding the trained model $h$. However, a malicious aggregator could also (i) sample a fresh dataset $S_2'$, (ii) encode it obtaining $E(S_2')$, and (iii) use it along with $E(S_1)$ to obtain another model on the underlying dataset $S_1$ and $S_2'$. In fact, a malicious aggregator could learn arbitrarily many different new models on $S_1$. While a bit innocuous looking, such a simple attack can be quite problematic in general and is prevented by the standard notion of physical privacy for MPC protocols. But protocols that achieve this very strong notion of privacy \emph{inherently} require more than one round of interaction.

This means that, to analyze the privacy of an instance encoding mechanism, we cannot follow the path of first proving physical privacy and then analyzing functional privacy. Instead, in order to understand the privacy of instance encoding protocols we must analyze the leakage of each message sent by each party individually. On the positive side, if we can show that this leakage is small, then we do not need to worry about anything else as this encoding is the only information that each party reveals about their data. Also, presence of malicious parties will not change the leakage as each party performs locally and independent of all other parties. In the next subsection we propose leakage measurement approaches for instance encoding and then in the next section, we aim at understanding the minimum possible leakage of an instance encoding based on our proposed leakage formulation.


\subsubsection{Privacy Definitions for Instance Encoding}
\paragcompact{Private learning through instance encoding.} We now define a minimal  privacy notion for  (encoding-based)  learning protocols  $(E,L)$ for a learning problem $(\cD, H, C)$   where $E$ is a dataset encoding scheme 
and $L$ is a learning algorithm that works on encoded datasets. The definition is of the ``cryptographic'' indistinguishability flavor.

For privacy, we define two attack models both of which are privacy notions for the encoding itself --- meaning that the privacy requires the encoding  to hide the sensitive information.  If the encoding can hide the input so that it is hard to distinguish from other inputs, or at least hard to recover, then the model trained on encoded instances would also be private by standard post-processing arguments. We stress that both notions below can be studied for dataset encodings and the special case of instance encodings (where  $E$ is an instance encoding).


\begin{definition}[Instance distinguishing  attacks for dataset encoding mechanisms] \label{def:data_dist} The adversary $A$ selects a concept function $c$, and instances $\set{x_0,x_1,\dots,x_n}$ such that $c(x_0)=c(x_1)$ and sends them to the challenger.
The challenger shapes sets $S = \set{(x_i,c(x_i)) \mid 2 \leq i \leq m}$,   $S_0 = \set{(x_0,c(x_0))} \cup S$ and $S_1 = \set{(x_1,c(x_1))} \cup S$.
Then the challenger samples a random  bit $b\gets \set{0,1}$, encodes $S_b$ to get $\tilde{S}\gets E(S_b)$, and sends $\tilde{S}$ to the adversary. Given $\tilde{S}$ the adversary announces its guess $b'$ (about $b$). The advantage of the adversary (against $c$) is defined as $p-1/2$ where $p$ is the probability that $b=b'$.
\remove{
\begin{align*}
    &\Advan(A, c, n)=\\
    &\quad\bigg|\Pr_{S\gets D_c^{n-1}}\Big[A\Big(E\big(S \cup \set{(x_1,c(x_1))}\big)\Big)=1\Big] \\
    &\quad - \Pr_{S\gets D_c^{n-1}}\Big[A\Big(E\big(S \cup \set{(x_0,c(x_0))}\big)\Big)=1\Big]\bigg|.
\end{align*}
}
\end{definition}
Note that this definition captures a  weaker notion compared to differential privacy, as the sets are both consistent with the same concept function, and even where they differ the two points still have the same label. In fact, when we prove limits of privacy under Definition \ref{def:data_dist}, the adversary only states the distribution of the instances in the set $S$ wothout picking them!  

Next, we consider a slightly weaker distinguishing game for the special case of instance encodings where the attacker is given an encoding of just one sample. This makes the task of distinguishing potentially easier for the  attacker. This setting is inspired by the InstaHide framework, but it is more general.
\begin{definition}[Instance distinguishing attacks for instance encoding mechanisms] \label{def:instance_dist}
This security game is defined for an instance encoding mechanisms $E=(E_X,E_Y)$. The adversary $A$ selects a distribution $\cD$, a concept function $c$, and two instances $x_0$ and $x_1$ such that $c(x_0)=c(x_1)$. Then the encoder samples $S = (x_2, x_3, \dots, x_n) \gets D^{n-1}$ and a bit $b\gets \set{0,1}$ and encodes  $ E^{1}_X(x_b, x_2,\dots,x_n)$ to get $\tilde{x}$. Given $\tilde{x}$ the adversary must decide whether $b=0$ or $b=1$ by outputting $b'$. The advantage of the adversary (against $c$) is defined as $p-1/2$ where $p$ is the probability that $b=b'$.
\remove{
\begin{align*}
    \Advan(A, c, n)&=\bigg|\Pr_{S\gets D^{n-1}}\Big[A\Big(E^{1}_X\big(x_1,S\big)\Big)=1\Big] \\
    & - \Pr_{S\gets D^{n-1}}\Big[A\Big(E^{1}_X\big(x_0, S\big)\Big)=1\Big]\bigg|.
\end{align*}
}
\end{definition}

Finally, we consider a weak form of privacy that prevents an adversary from \emph{recovering} parts of an input given its encoding.

\begin{definition}[Instance recovering attacks] A dataset $S=\set{(x_1,y_1),\dots,(x_n,y_n)}$ is encoded to $(\tilde{X},\tilde{Y}) \gets E(S)$ and given to the adversary. The goal of the adversary is to find a $x^*$ such that $d(x^*,x_i)\leq \gamma$, for some $i\in [n]$ under some (context-dependent) metric $d(\cdot,\cdot)$. 
\end{definition}


Distinguishing attacks are harder to defend against. In the following section, we give a barrier against achieving privacy against distinguishing attacks. Note that our result does not rule out the possibility of privacy against instance-recovering attacks. Indeed, to rule out such attacks, one has to first choose a natural metric (e.g., based on some $\ell_p$ norm), which is context dependent. In contrast, our results in Section~\ref{sec:impossibilities} are general. 

\section{Barriers for Privacy with Instance Encoding}

In this section, we present distinguishing attacks against learning protocols equipped with an instance/dataset encoder.
We first prove a theorem in the most general setting. Namely, we consider general dataset encoding mechanisms and show the existence of dataset distinguishing attacks.


Due to space limitations, all proofs are moved to Appendix~\ref{sec:proofs}.

\begin{theorem}[Formal statement of Theorem \ref{thm:Inf1}]\label{thm:dataset_encoding}
Let $c_1$ and $c_2$ be two distinct\footnote{By distinct we mean $c_2$ is not identical to  $c_1$ or $1-c_1$.} and non-constant concept functions for inputs $X$ and labels $\bits$. 
Let $D$ be a distribution over $X$ such that $\Ex[c_2(D)]=0.5$ and $\Ex[c_1(D)]=0.5$.
 If a protocol $(E,L)$ can achieve $(0.51,\delta)$-accuracy on plain data over both of $D_{c_1}$ and $D_{1-c_1}$, then there is an dataset distinguishing adversary for $(E,L)$ against either $c_1$, $1-c_1$ or $c_2$ with advantage at least ${(0.99-2\delta(n))}/{3n}$ (according to Definition \ref{def:data_dist}), where $n$ is the size of the dataset.  Moreover, the running time of this adversary is essentially the running time of $L$.   
\end{theorem}

\paragraph{Discussion.} Theorem~\ref{thm:dataset_encoding} gives a \emph{distinguishing} attack of advantage $\Omega(1/n)$. Since the two datasets used by the adversary in the proof are neighbors (i.e., differ in one point), this also implies a lower bound on the sample complexity of differentially private learners (based on the level $\eps$ of differential privacy). This result further shows that none of the restrictions on the adversary (as stated in Theorem~\ref{thm:dataset_encoding}) can limit the adversary's distinguishing advantage (or the corresponding $\eps$ in a candidate differentially private scheme) to $o(1/n)$. In fact, the proof of Theorem~\ref{thm:dataset_encoding} shows something stronger: the adversary will \emph{not} pick the core  set $S$ that is shared between $S_0,S_1$, but rather that set is \emph{sampled} from a distribution chosen by the adversary. Finally, we note that the complexity of the concept functions in Theorem \ref{thm:dataset_encoding} cannot be reduced to having only one concept function. That is because, if $C=\set{c}$, the learner can basically ignore the data and just output a canonical representation of $c$, leading to a perfectly private scheme.

The impossibility result above does not consider the scenario where the encoding mechanism can get some auxiliary information about the concept function. Specifically, we assume that the only information that the encoder obtains from the underlying concept is through the dataset. In fact, if the concept function $c$ was directly known to the encoding mechanism, it could simply output a description of $c$ and hide the input data.


Next, we consider the setting of local instance encodings that are applied independently to each training sample (i.e., a $1$-local encoding). Our first result applies to learning tasks with a rich class of concepts, as formalized hereafter.

\begin{definition}[Rich concept class] \label{def:rich}
For concept class $C$ and given parameters $m \in \N,\gamma \in \R^+$, we say that the concept class $C$ is $(m,\gamma)$-rich with respect to distribution $\cD$, if there exists a vector $F=(c_1,\dots, c_m)$, $c_i \in C$ with the following  property: For any configuration $f\in\bits^{|F|}$
$$\Pr_{x\gets D}\left[ \gamma\leq \frac{ |F(x) - f|} {|F|} \right] \geq 0.99 $$
\end{definition}

It is easy to see if that if the concepts $c_1,\dots,c_m\in D$ are all balanced and orthogonal (the probability of every output $f \in \bits^m$ to be produced by them over a random $x \gets D$ is $2^{-m}$), then by standard Chernoff-type arguments, the $(m,\gamma)$-richness property holds for any constant $\gamma > 0$ and sufficiently large $m$. (The balanced and orthogonal setting was used as a special case when stating Theorem~\ref{thm:rich} informally in Section~\ref{ssec:overview}). We now state the formal version of our result.

\begin{theorem}[Formal statement of Theorem \ref{thm:Inf2}: Barrier for privacy with instance encoding on a rich concept class] \label{thm:rich} Consider a learning problem $(\cD, C, H)$ where $C,H\subset \set{0,1}^X$  and where $C$ is $(m,\gamma)$-rich according to Definition \ref{def:rich}.
If a learning protocol with encoding $(E,L)$ has encoded accuracy $(\epsilon,\delta)$ on this problem. Then, for any $c\in C$ there is an instance distinguishing attack (according to Definition \ref{def:instance_dist}) $A^{L(\cdot), E(\cdot), F(\cdot), \cD}$ that has oracle access to $E(\cdot)$, $L, F$ and a sampler for $\cD$ and gets advantage $0.99-\frac{\epsilon(n)}{\gamma}$ against $E$ according to Definition \ref{def:instance_dist}. The expected running time of this adversary is $O(\frac{m}{1-\delta(n)})$. Moreover, the attacker's samples $(x_0,x_1)$ are sampled jointly from the same distribution $D$ conditioned on labels being the same. 
\end{theorem}

\paragraph{Discussion.} The idea behind the proof of Theorem~\ref{thm:rich} is that if the learned model has non-trivial \emph{encoded accuracy} (i.e., we can predict the label of an instance from its encoding), then this leakage already implies a (possibly weak) distinguishing attack between encodings. To amplify the attack's distinguishing power, we leverage the fact that we can learn multiple concept functions from the class $C$ using the \emph{same} encodings.

Theorem \ref{thm:rich} shows a barrier against achieving both indistinguishability privacy and encoding accuracy on a rich class of concept functions. Theorem \ref{thm:single_concept} below shows a barrier for the orthogonal case where the encoding can depend on the concept function itself (e.g., if there is just one concept to learn). In particular, for the following theorem, we do not require the protocol to work for multiple concept functions and it can be tailored to a specific concept function. The same argument we use to prove Theorem~\ref{thm:rich} above will not work anymore, as the protocol might use an entirely different encoding for different tasks and a classifier trained for one task will not be a good distinguisher for the encodings of other tasks.

Note that, in the extreme case, the encoding could completely depend on the concept function $c\in C$. For example, imagine an encoding algorithm that maps each instance to its correct label. This encoding is perfectly secure against the distinguishing attacks of Definition \ref{def:instance_dist}. This encoding can also achieve $100\%$ accuracy if an identity classifier is applied to it. Therefore, there is no privacy versus accuracy trade-off for this case. However, we can still prove some barriers against privacy if we assume that learning a perfectly correct classifier is hard. Bellow, we show that if an encoding achieves both reasonable privacy and accuracy, then it is possible to efficiently extract an almost-perfect classifier from it. 
%

\begin{theorem}[Barriers for privacy with instance encoding on a single concept]\label{thm:single_concept} Consider a learning problem $(\cD, C, H)$ where
 $H,C\subset \set{0,1}^X$. Also assume that for a concept $c\in C$, $\Pr[c(\cD)=1]=0.5$. Consider an efficient learning protocol with decomposable instance encoding $(E,L)$ that has \emph{balanced} $(\epsilon,\delta)$ accuracy on encoding for $c$ and according to distribution $\cD$. Then, for any $\tau\in [0,1]$, one of the following is correct:
 \begin{itemize}
     \item{\textbf{Lack of privacy:}} There is an efficient attack with oracle access to $L, E$ and $\cD_{c}$, that runs in expected time $O(\nicefrac{m}{\delta(m)} +\nicefrac{m}{\tau^2})$ and has average advantage (according to Definition \ref{def:instance_dist}) at least $\frac{1}{2} - \epsilon(m) -\tau$ in winning in the instance  distinguishing game (Definition~\ref{def:instance_dist}).
     \item{\textbf{Very high accuracy:}} There is an efficient learning protocol $(L',E')$ that learns this problem (privately) using $m$ samples and outputs a classifier $h'$ (with running time $O(m/\tau^3)$) that has accuracy at least $1-\tau$.
 \end{itemize} 
\end{theorem}

This theorem shows that if an encoding function makes all examples of a class indistinguishable from each other, then that encoding must contain almost all the information that a perfect classifier has (and this information can be extracted efficiently). 
This shows a barrier against privately learning tasks that have a lower bound on their sample complexity. For example, if we know that a problem $(\cD,H,C)$ is not learnable with accuracy more that $95\%$, then it is not possible to learn it privately with accuracy more than around $50\%$ on the encoded data.


\newcommand{\goodset}{\mathrm{G}}

\label{sec:impossibilities}


\section{An Attack on InstaHide}
\label{sec:instahide_attack}

\label{sec:attack}

The above formal analysis applies to \emph{any} encoding-based scheme.
To make our analysis concrete, we now introduce a reconstruction attack on InstaHide \cite{huang2020instahide},
an instance-encoding scheme published at ICML 2020
and awarded the 2nd place 2020 Bell Labs Prize.
Given access to a set of encoded images, this attack recovers the original images that were used to generate the encoding.

\subsection{Background}

InstaHide proceeds as follows.
First, gather a large public dataset $p \in P$, e.g., of arbitrary images from the Internet.
Then, generate the encoded dataset  $(e,z) \in E$ (representing encoded images $e$ with encoded labels $z$)
by assigning
\begin{align*}
    E \gets \big\{ & (XMix(\{x_i, x_j\}, p, \lambda), YMix(y_i, y_j, \lambda) \\
    & \colon ((x_i, y_i), (x_j, y_j)) \in X, p \subset P, |p| = k-2 \big\}. 
\end{align*}

The core algorithms in InstaHide, $XMix$ and $YMix$, are defined as follows.
\[ XMix(x, p, \lambda) = \sigma \circ \left(\sum_{i=1}^2 x_{i} \lambda_{i} + \sum_{i=3}^k p_{i-2} \lambda_{i}\right) \]
with $\lambda$ chosen uniformly at random such that $\sum_i \lambda_i = 1$;
the mask $\sigma$ chosen uniformly at random from $\sigma \in \{-1,1\}^d$,
and where $a \circ b$ denotes element-wise multiplication.
The function $YMix$ is much simpler and given by
\[YMix(y_i, y_j, \lambda) = y_i \lambda_1 + y_j \lambda_2\]
with addition taken component-wise across one-hot labels.
The size of the encoded dataset is determined by the
encoding multiple $N = |E|/|X|$, with each instance being encoded $N$ times.
In practice, this multiple is equal to the number of training epochs (e.g., 50 or 100).
The authors argue InstaHide is secure for $k \ge 4$, with the strongest version at $k=6$ (e.g., the InstaHide
Challenge released by the authors uses $k=6$).

We make use of some additional notation.
Let $\phi : E \to (|X| \times |X|)$ represent the mapping from the encoded images to original private images. 
By $\phi(e_i) = (j,k)$ we mean that encoded image $e_i$ is built out of
the original images $x_j$ and $x_k$.
Similarly, let $\phi^{-1}$ be the inverse so that $\phi^{-1} : X \to 2^{|E|}$, for example
$i \in \phi^{-1}(x_j)$ and $i \in \phi^{-1}(x_k)$.
Note that while $\phi$ maps one $x \in X$ to \emph{exactly} two $e_1, e_2 \in E$, the inverse
$\phi^{-1}$ maps one $x \in X$ to approximately $2N$ encoded images $e \in E$.

\subsection{Attack Overview}

\begin{figure}[t]
    \centering
    \includegraphics[width=\columnwidth]{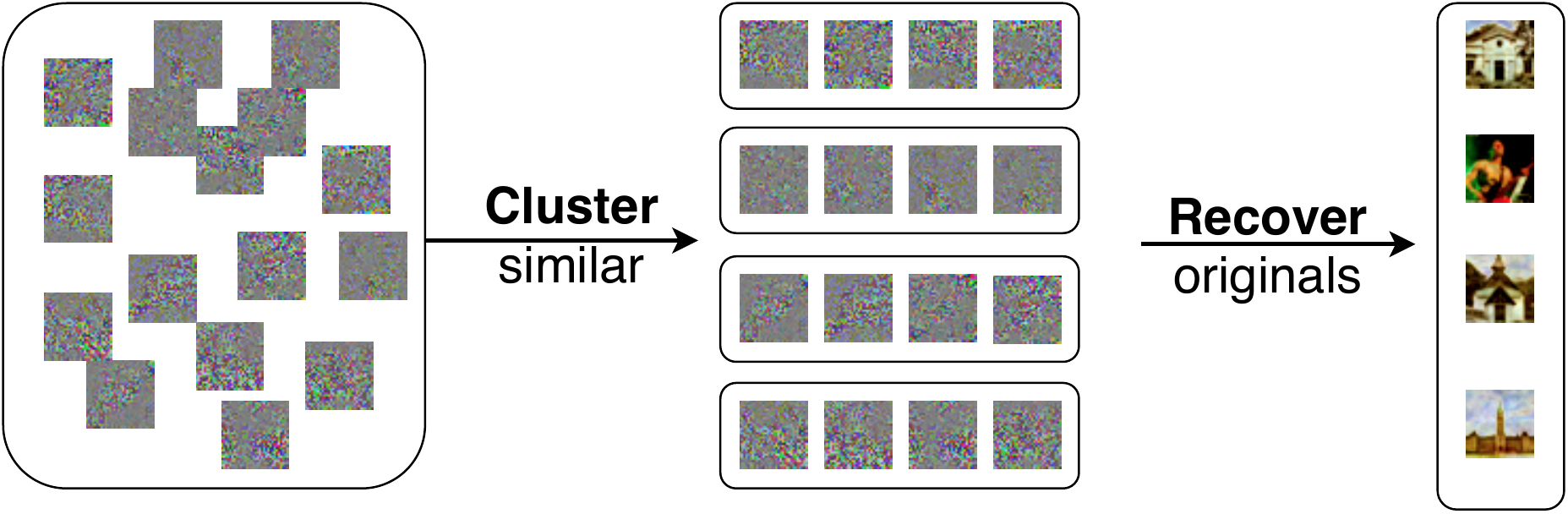}
    \caption{{Our attack process on InstaHide encodings.} Given the encoded dataset,
    we cluster together images generated from the same original source
    image and then from these sets ``decrypt'' them to the original sources.}
    \label{fig:attack_process}
\end{figure}

We break InstaHide's privacy through an attack that consists of three stages:
\begin{enumerate}
\item \textbf{Remove instance hiding:} Replace the encoded dataset by
\[E \gets \{abs(e)\,\colon\,e \in E\}\]
which nullifies the sign flipping step in $XMix$.

\item \textbf{Cluster encoded dataset:} Given these absolute-value images, 
we recover the mapping $\phi$ that determines which
original images were used to generate each encoded image.

We achieve this by training a neural network to detect when two encodings
were generated from the same original image.
This lets us build a graph of pairwise similarity between encodings,
from which we can extract one clique per original image with the vertices in this clique
corresponding to the encodings generated from that original image.

\item \textbf{Recover original images:} Then, given the encoded images and the mapping $\phi$, we
recover (an approximation of) the original labeled images $X$.

This step involves solving an under-determined (nonlinear) system of equations
via gradient descent. Because the system is under-determined, it is provably impossible
to recover the original images pixel-perfect, however this does not prevent reconstructions
that have high similarity to the original images both qualitatively and quantitatively.
\end{enumerate}

We release the source code of our attack as a utility that can be used to break the privacy
of arbitrary InstaHide encoded images. 
As we will show, our attack is hyperparameter free (except for the sizes of the images)
and the one configuration we release breaks the privacy of InstaHide on
CIFAR-10, CIFAR-100, and the InstaHide challenge \cite{challenge}.


\subsection{Clustering}
\label{subsec:clustering}

The purpose of the clustering stage is to recover
$\phi$, the function that maps
original source images to encoded images.
Because each encoded image has two original images that were used to
generate it, our goal is to recover $|X|$ sets $S_i$ of encoded images, where each
set has size about $|S_i| \approx 2N$. 
%
At the end of this step, we will know which encoded images were generated using each original image $x_i$.

This stage follows five steps.
\begin{enumerate}
\item Create a pairwise similarity function $sim(e_i, e_j) \to [0,1]$ so that
$sim$ is high if $e_i$ and $e_j$ share at least one source image and low otherwise.
\item Construct the complete weighted similarity graph $G$ that represents the all-pairs similarity.
\item Find sets $\{S_j\}_{j=1}^{|X|}$ by finding densely connected cliques.
\item Construct a new bipartite graph that maps the similarity between each encoded $e_i$ and the nearest set $S_j$.
\item Assign each encoded image $e_i$ to two sets $S_j$, and assign each set $|N|$ encoded images, minimizing total cost.
\end{enumerate}

\subsubsection{Learning a Similarity Function}
\label{sec:first}

Our first step of the attack constructs a similarity function $sim$ that
determines if two images $e_i$ and $e_j$ were generated using at least one
shared original image.

\paragraph{Inputs:} 
The public dataset $P$.

\paragraph{Outputs:}
The function $sim$, so that $sim(e_i, e_j)$ is (usually)
1 if $\phi(e_i) \cap \phi(e_j) \ne \varnothing$ and 0 otherwise.

\paragraph{Method:}
We train a neural network to approximate this similarity function $sim$.
We create a large training dataset with examples
of pairs of images encoded together and not.
This neural network receives the two inputs $e_i$ and $e_j$
stacked on the channel dimension (so, concretely, for $32 \times 32 \times 3$
color images the input to the neural network is $32 \times 32 \times 6$).
The neural network outputs a single scalar $y \in \mathcal{R}$ and we
assign a standard sigmoid loss so that $y>0$ when the two images
share an original image and $y<0$ otherwise.

We train a single neural network to be used for all attacks in this paper.
We use a Wide ResNet-28 trained with Adam with a learning rate of $0.1$ and a
weight decay factor of $5\cdot 10^{-4}$ for $10^6$ steps.
We use a 32x32 downsampling of ImageNet as the public dataset following
the process described in~\cite{huang2020instahide},
and the CIFAR-10, CIFAR-100, and STL-10 training images as the private
images.
We augment the training process with standard flips and shifts.
The final trained model reaches $91\%$ accuracy on a held-out validation set.

\subsubsection{Constructing the Similarity Graph}

\paragraph{Inputs:} 
The encoded images $E$,
and the similarity function $sim$ from the prior subsection.

\paragraph{Outputs:}
A complete weighted similarity graph $G$ that has an edge between
each encoded image $e_i$ and $e_j$ with weight equal to $sim(e_i,e_j)$.

\paragraph{Method:}
This step is trivial.
We evaluate the neural network on all $|E|^2$ pairs of images.
For modestly sized encoded datasets this process is efficient, for example
on the $5{,}000$ image contest dataset this step finishes in $10$ minutes.

\subsubsection{Identifying Densely Connected Cliques}

\paragraph{Inputs:} 
The weighted graph $G$ from the prior subsection.

\paragraph{Outputs:}
A coloring of the vertices into $|X|$ non-overlapping subsets 
$\mathcal{S} = \{S^{(i)}\}_{i=1}^{|X|}$ that approximately maximizes
\[\sum_{S \in \mathcal{S}} \sum_{e_i \in S, e_j \in S} \text{weight}(e_i,e_j). \]
In an ideal reconstruction, we would have that
\[\bigg| \bigcap_{e \in S^{(i)}} \phi(e)\bigg| = 1  
\hspace{4em} 
\forall i \neq j:\ \bigg(\bigcap_{e \in S^{(i)} \cup S^{(j)}} \phi(e)\bigg) = \varnothing 
\]
That is, each subset contains encodings that share exactly one source image (the \emph{representative} of this subset). Moreover, no two subsets have the same representative.

\paragraph{Method:}
The purpose of this algorithm is to create $|X|$ clusters, one for each
original image in the dataset.
Note that each encoded image is actually created from $2$ different original
images; however, for now, we will simply assign each encoded image to just one
original image.
That allows this step to be a simpler problem of ``coloring'' this graph
with $|X|$ different colors minimizing cost.

We develop a greedy algorithm to approximately solve this problem.
The core of our algorithm is a recursive loop that iteratively selects the
next best encoded image to add to an existing set using the update rule
\[insert(S) = S \cup \bigg\{ \mathop{\text{arg max}}_{e \in E} \sum_{u \in S^{(i)}} \text{weight}(e,u) \bigg\}. \]
That is, we greedily add the closest example that has the highest weight
when considering those examples that are already in the set.
Then we define
\[create(S, M) = \underbrace{insert(insert(\dots(insert}_{\text{repeated $M$ times}}(S))\dots)\]

This lets us compute the sets $T^{(i)} = create(\{e_i\}, M)$
for each encoding $e_i \in E$.
To choose the integer $M$ we select a constant $M < N/2$ (we found that setting $M = N/4$ works in practice).
At this point, we should expect that there are $|X|$ distinct sets among the collection of sets $\{T^{(i)}\}_{i=1}^{|E|}$. 

\paragraph{Justification:}
If each step up until this point was perfect (i.e., if the similarity neural network
returned $1$ if and only if two encoded images were generated from the same source image)
then with probability almost $1$ we would expect exactly $|X|$ distinct sets: one for each original image.
That is, formally, we can inductively prove that
$\left|\bigcap_{s \in T^{(i)}} \phi(s)\right| > 0$
(and with overwhelming probability this intersection
contains exactly one element).
To see that this is the case, when we start with a set containing a single element $\{e_i\}$ and call
$\{e_i, e_j\} \gets insert(\{e_i\})$ we are guaranteed to have that 
$e_i$ and $e_j$ share at least one original image $x$ (formally, 
$|\phi(e_i) \cap \phi(e_j)| > 0$).
With probability ${\frac{1} {|X|}}$ we should expect
$|\phi(e_i) \cap \phi(e_j)| = 1$ because each encoded image is constructed by
pairing together two original images at random, and so the probability that two encoded
images share both original images given that at least one is identical is ${\frac{1}  {|X|}}$.
The inductive case is identical.

Importantly, if $\phi(e_i) = (x_a, x_b)$, then both of the original images $x_a$ and $x_b$ have equal probability of also being part of some other encoding $e_j$.
Thus, consider each encoded image $e$ that is generated
using the image $x_b$.
The probability that
\[x_b \not\in \bigcup_{e \in \phi^{-1}(x_b)} \left( \bigcap_{x \in create(e)} \phi(x) \right) \]
is exactly $1/2^N$,
as this happens only if each call to $create(e)$ creates a set based around
the \emph{other} private image used to generate that encoding $e$.
Thus, with $N=100$ as we have in our experiments, we can discount this ever happening.
This allows us to conclude that we will have $|X|$ sets.

Unfortunately the prior steps are not perfect.
As a result, it is possible to have $\epsilon < | T^{(i)} \cap T^{(j)} | < N$
for $\epsilon$ an integer greater than zero.
We can still solve this problem approximately, however.
Given the $|E|$ sets, we want to cluster them into $|X|$ clusters-of-sets
where we maximize the similarity of the sets in individual clusters.
To do this, we perform k-means clustering on these sets (with $k=|X|$), where the
distance between sets is defined as $d(s, t) = \frac{|s \cap t|}{|s \cup t|}$.
We run this to cluster the sets into $|X|$ different clusters and then
choose one representative (arbitrarily) from each cluster to form
the sets $S^{(i)}$.

\subsubsection{Computing Similarity Between Encodings and Cliques}

\paragraph{Inputs:} 
The encoded images $E$, the $|X|$ (near-)cliques $\mathcal{S}$.

\paragraph{Outputs:}
A new graph $G'$ that computes the distance from any encoded image $e \in E$
to each of the other sets $\mathcal{S}$.

\paragraph{Method:}
The simplest strategy just computes the average
$\sum_{v \in S} \text{weight}(e, v)$ for each $S \in \mathcal{S}$.

We can do better, though.
This similarity graph was constructed with a neural network that receives \emph{two} encoded images and tests whether they share
an original image.
Our problem is now easier: we have $|S^{(i)}|$ encoded images,
all of which (probably) belong to the same original image $x$, and
we want to test if an encoded image $e$ also belongs
to the same original image $x$.

We thus train a new similarity neural network to
return $1$ if an encoded image $e$ shares the same original
image as a set of examples $\{e_1,e_2,e_3,\dots\}$.
We find experimentally that we reach diminishing returns 
once we provide the neural network with more than $4$ examples.
This new task is easier for the network to solve. 
By having $4$ examples of what the original image looks like, it is
easier for the model to learn to predict if a $5$th image uses
a similar base image.
In practice, this new neural network increases the prediction
accuracy from $91\%$ to $96\%$ (reducing the error rate by a factor of $2$).

To construct the similarity graph $G'$ we choose four images in each set $S$ at random.
Then, we compute the distance from each $e \in E$ to 
the four representatives from each set, 
giving us a bipartite graph connecting the $|X|$ sets to the $|E|$ examples.

\subsection{Assigning an Encoded Image to an Original Image}

\paragraph{Inputs:} 
The new similarity graph $G'$.

\paragraph{Outputs:}
A mapping $\phi'$ that maps encoded images to original images.
Ideally, we will have that $\phi' = \phi$.

\paragraph{Method:}
We can solve the final assignment problem
with a single call to min cost max flow \cite{edmonds1972theoretical}.
We construct a source node with a supply of $2 |X|$, and a
sink node with a supply of $-2 |X|$.
Then, we connect the source to each set $S$ with capacity
$|N|$, each set $S$ to each example $e_i$ with capacity $1$,
and each example $e$ to the sink with capacity $2$.
The min cost max flow assignment will therefore assign each
example $e_i$ to exactly two sets $S$, and assign each set
to exactly $|N|$ distinct examples $e_i$, exactly satisfying
the constraints specified for $\phi$.
This gives us the mapping function $\phi'$.

The fact that each encoded image correspond to exactly two
original images, and each set contains exactly $N$ encoded
images, is built into the design of the InstaHide algorithm:
instead of randomly choosing two images to pair together to
form each encoded image, InstaHide generates two random
permutations of the original images $p^{(1)}$, $p^{(2)}$ and then
pairs together the elements in this sequence, so 
$e_1$ is generated from $p^{(1)}_1$ and $p^{(2)}_1$,
through to $e_N$ generated from  $p^{(1)}_N$ and $p^{(2)}_N$.
A new permutation is then generated, and the process repeats.

If InstaHide instead randomly selected sets of size approximately
(but not exactly) $|N|$ our attacks would remain effective;
it would require a slightly modified scheme
but preliminary experiments suggest that attack success rate
remains unchanged.


\subsection{Recovery of the Original Images}

Given the resulting images pairings $\phi'$, 
we must now reconstruct the actual values of the original images.

\subsubsection{A Simple Proof of Concept}

At this stage, we can gather all encoded images $\{e_{x_i}\}$ that 
include the same original image $x$ by inverting the recovered mapping $\phi'$.
Then, by computing the pixel-wise mean after taking the absolute value
$\tilde{x_i} = \text{mean}_{e \in S^{(i)}} abs(e)$ we obtain an
approximation of the absolute value of the original images.

Why does this work? By taking the absolute value, we remove
the pixel-flipping information-hiding induced by multiplication with 
$\sigma$.
Then, by taking the pixel-wise mean we ``average out''
the noise from all of the other images that are mixed up with this one image,
which gives us just the signal.

This recovers visually recognizable images, but
(a) we have lost the sign information, and more importantly
(b) we introduce a large amount of visual noise to the resulting images.

\subsubsection{Recovering the Mix-Up Values $\lambda$}
In order to do better, we will first need to recover not only
$\sigma$ but also the mix-up values of $\lambda$ used.
Fortunately, this step is (almost) trivial. The \emph{unordered} values
of $\lambda$ are provided to the adversary by the InstaHide algorithm
in the form of the labels $z$---each label in InstaHide is also mixed up
directly.

As a result, we can (almost directly) read off the coefficients of $\lambda$ with one
exception: if InstaHide mixes up two images of the same class, then
we obtain a single label with value $l = \lambda_i + \lambda_j$.
Because it is impossible to disentangle these values, we simply
guess $\lambda_i = \lambda_j = l/2$.

\subsubsection{Recovering Original Images Assuming no Sign Flipping}
\label{sec:partialthing}

Given this additional information of $\lambda$ we show how to improve the recovery of the original images. To simplify exposition, we begin by assuming that InstaHide does not perform
any pixel-flipping by multiplying images with $\{-1,1\}^d$.

\paragraph{Inputs:}
The encoded images $E$ (without pixel flipping), the mapping $\phi'$, and the values of $\lambda$.

\paragraph{Outputs:} The (near) original images $X$.

\paragraph{Method:}
This attack is straightforward least squares.
Let $A$ be a  $|X| \times d$ unknown matrix (if solved for correctly, with rows corresponding to images $x$).
Let $B$ be a  $|E| \times d$ known matrix with rows corresponding to images $e$.
\[ B = \left[ \begin{array}{llll} e_1 & e_2 & \dots & e_{|E|} \end{array} \right]^T \]

Then finally let let $M$ be a sparse $|E| \times |X|$ dimensional matrix that is
zero almost everywhere except when $\phi(i) = (j,k)$ where
\[ M_{i,j} = \lambda_{e_i,1} \quad\text{and}\quad M_{i,k} = \lambda_{e_i,2}. \]

Therefore if $A$ was correct then we would have that
\[ M \cdot A = B + \sigma. \]
where $\sigma$ is the noise component for the public images
(factored out).
Therefore we can ``just'' solve for the equation
\begin{align*}
    A = M^{-1} (B + \sigma) = M^{-1} \cdot B + M^{-1} \sigma \approx M^{-1} \cdot B
\end{align*}
assuming that $\sigma$ is distributed normally.
The reason this holds true is that if $\sigma$ is symmetric about zero,
then the expected mean value of $M^{-1} \sigma \approx \vec{0}$.

Put differently, what we're effectively
doing is minimizing the ``unexplained variance'' by minimizing
\begin{align}
    \mathop{\text{arg min}}_{A' \in [-1,1]^{|X| \times d}} & \lVert B - M \cdot A' \rVert_2^2
    \label{eqn:min}
\end{align}
because the true solution to this equation would give
\[  \lVert B - M \cdot A \rVert_2^2 = \lVert (M \cdot A + \sigma) - M \cdot A \rVert_2^2 = \lVert \sigma \rVert_2^2. \]
and so this approach is well justified as long as minimizing $\sigma^2$ is the correct objective---and it
is for isotropic Gaussian noise.

\subsubsection{Recovering Original Images on Full InstaHide}

It is more difficult to solve the above equation if we mask the images by multiplying
with a random $\{-1,1\}^d$ vector.
However, we can still rely on the same intuition as before.

Solving Equation~\ref{eqn:min} is the same as solving the formulation
\begin{align}
    \label{eqn:min2}
    \mathop{\text{arg min}}_{A' \in [-1,1]^{|X| \times d}} & \lVert \sigma \rVert_2^2 \\
    \text{such that} \,\,\, & M \cdot A' + \sigma = B. \nonumber
\end{align}
This modified formulation is identical, but while Equation~\ref{eqn:min} will
not generalize to the full InstaHide Equation~\ref{eqn:min2} will.
To do this, we modify the minimization to instead solve
\begin{align}
    \mathop{\text{arg min}}_{A' \in [-1,1]^{|X| \times d}} & \lVert \sigma \rVert_2^2 \\
   \text{such that} \,\,\, & M \cdot abs(A') + \sigma = abs(B) \nonumber
\end{align}
where $abs$ is taken component-wise on the matrix.


We search for $A'$ via gradient descent.
Given an attempted solution $A'$ we can use the constraint
$M \cdot abs(A') + \sigma = abs(B)$ to solve for $\sigma$,
which then lets us compute the objective 
$\lVert \sigma \rVert_2^2$.
There is one complication here: given a matrix $A'$, there are
multiple values $\sigma$ which satisfy the above constraint.
Fortunately, because we know that it is our objective to minimize $\lVert \sigma \rVert_2^2$
we can greedily choose each entry $\sigma_{ij}$ as the smaller of the two candidates.
Along with being much more computationally efficient,
this approach has the benefit that we can solve the $\ell_2$ norm minimization as well.

\subsection{Adjusting Color Saturation Levels}

Given the recovered original images, we
repair their saturation levels to better reflect the distribution of natural images.

\paragraph{Inputs:} The reconstructed images $X$.

\paragraph{Outputs:} The color-adjusted images $X_{fixed}$.

\paragraph{Method:} 
We find that while the reconstructions are of high quality, saturation curves are
misaligned between the original and the reconstructed inputs.
Manual adjustment of these curves is effective, but we can develop an automated approach.

We train a tiny (73 total parameter) neural network for this task.
The network receives as input a single pixel (3 RGB colors), has a 10-neuron
hidden state, and then outputs a single pixel with the new color values.
To train this model, we create a new challenge using our own images, run the full attack
up to this point, and record the reconstructed images along with the original images.
Then, we create a training dataset mapping the reconstructed pixel values onto the original
pixel values.
We train this model for one epoch on $100,000$ training examples, and then apply it on
the final images for each of our attacks.

\begin{figure}
    \centering
    \footnotesize
    \begin{subfigure}[b]{\columnwidth}
         \begin{tabular}{@{}p{0.7in}@{}}
              Original  \\
              \vspace{1em} \\
              Reconstructed \\
              \vspace{3em}
         \end{tabular}
         \includegraphics[width=0.78\textwidth]{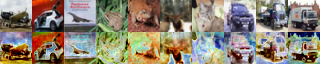}
         \vspace{-4em}
         \caption{CIFAR-10}
         \label{fig:break10}
     \end{subfigure}
    \vspace{-.5em}
    
    \begin{subfigure}[b]{\columnwidth}
         \begin{tabular}{@{}p{0.7in}@{}}
              Original  \\
              \vspace{1em} \\
              Reconstructed \\
              \vspace{3em}
         \end{tabular}
         \includegraphics[width=0.78\textwidth]{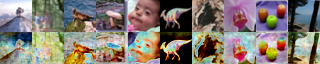}
         \vspace{-4em}
         \caption{CIFAR-100}
         \label{fig:break100}
     \end{subfigure}
    \vspace{-1em}
    
    \caption{Our reconstruction attack on InstaHide evaluated on CIFAR-10 (a) and CIFAR-100 (b). 
    The top row of each subfigure contains 10 original images that were encoded,
    and the bottom row our reconstruction of that image.}
    \label{fig:works}
\end{figure}

\subsection{Results}

We evaluate our attack on the two datasets considered in the original paper:
CIFAR-10 and CIFAR-100.
We further evaluate our attack on an unknown dataset challenge released
by the authors consisting of $5{,}000$ encoded images from an unknown
distribution generated from $100$ original source images.

Because our attack is hyperparameter free and independent of any particular
dataset (as long as the images are the same size---fortunately, all datasets
considered are $32\times32$) we do not need to change any details to perform
the attack below.

We implement our attacks in JAX~\cite{jax2018github}, a numerically accelerated version of
NumPy with built in automatic differentiation.
We train our neural networks using Objax,\footnote{\url{https://github.com/google/objax}}.

\subsubsection{CIFAR-10 and CIFAR-100 Results}

\paragraph{Constructing the encoded dataset.}
We construct our own dataset by using the authors existing open
source code.\footnote{\url{https://github.com/Hazelsuko07/InstaHide}}
%

We take the first $100$ images in the test set, and then encode this to a dataset
of $5{,}000$ total encoded images using the $k=6$ InstaHide scheme described above.

Our attack is extremely effective across both of these datasets.
Figure~\ref{fig:works} shows the first $10$ images of the $100$ total images in
the dataset. The full $100$ examples are given in Appendix~\ref{apx:figures}.

Our attack is computationally efficient.
Computing the initial all-pairs distance takes two hours on one GPU,
finding the $|X|$ cliques takes 2 CPU-hours,
computing the $|X| \times |E|$ all-pairs graph takes $19$ minutes,
and the final recovery step takes $1$ minute.
In total, the attack took 2 GPU hours and 2 CPU hours.

\subsubsection{InstaHide Challenge Results}

The InstaHide Challenge \cite{challenge} 
was released by the InstaHide authors
as a public challenge to break InstaHide.
The authors use the strongest version of InstaHide and release
$5{,}000$ encoded images corresponding to $100$ private images.
Because only the encoded images are released, 
we do not have ground truth available and so can not visually compare our
results with the actual images.
Our attack takes under an hour to complete.

Figure~\ref{fig:instahide_break} shows ten of the original images that we recovered. The complete $100$ recovered images are given in Appendix~\ref{apx:figures}.

\subsection{Analysis of InstaHide's Security Parameters}
\label{sec:instahide_parameters}

The above reconstruction attack is fully general and breaks InstaHide under the defense settings described
by the authors and the released InstaHide challenge.
However, InstaHide has two ``security parameters'' that are claimed to
increase the security if set appropriately. 
Specifically, 
\begin{itemize}
    \item The total number of released images $|E|$ as controlled by the
    number of times the dataset is replicated $N$. A larger $N$ is necessary for accurate models (e.g., the InstaHide challenge sets $N=50$, but the security is claimed to increase with saller $N$.
    \item The MixUp-$k$ value controls the number ($k$) of original images
    used to form a single encoded image. The authors show that increasing $k$ decreases accuracy, but claim that increasing $k$ improves security.
\end{itemize}

We now introduce two attacks that show neither of these security parameters
significantly increase the actual security of InstaHide.
Even if $k>100$, InstaHide remains broken under the same attack as above,
and a new attack we develop can break InstaHide when $N=1$ epoch of data is released.

\subsubsection{Attacking InstaHide With a Single Encoding}
\label{ssec:attack_single_encoding}

\begin{figure}[t]
    \footnotesize
    \begin{subfigure}[b]{\columnwidth}
         \centering
         \begin{tabular}{@{}p{0.8in}@{}}
              Original Mixed  \\
              \vspace{0.5em} \\
              Recolored Mixed \\
              \vspace{3em}
         \end{tabular}
         \includegraphics[width=0.75\textwidth]{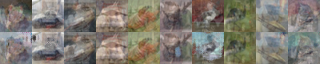}
         \vspace{-4em}
     \end{subfigure}
     
     \vspace{0.5em}
    
    \begin{subfigure}[b]{\columnwidth}
         \centering
         \begin{tabular}{@{}p{0.8in}@{}}
              Original  \\
              \vspace{0.5em} \\
              Reconstructed \\
              \vspace{3em}
         \end{tabular}
         \includegraphics[width=0.75\textwidth]{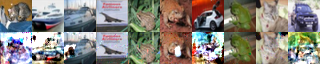}
         \vspace{-4em}
     \end{subfigure}

    \caption{Reconstruction attack on InstaHide evaluated on CIFAR-10 with a single encoding per private image. Our attack first trains a GAN to invert (i.e., ``re-color'') the absolute value of the mixed image (top). When the re-coloring succeeds, the private image is extracted near-perfectly by subtracting the public images with highest similarity to the mixture (bottom).}
    \label{fig:attack_single_encoding}
\end{figure}

Two core components of our attack on InstaHide, the clustering step and final image recovery step, exploit the fact that we have access to multiple random encodings of every private image. We now propose an alternative attack strategy that recovers private data \emph{given a single encoding} of each image. 

To achieve this stronger form of attack, we consider a stronger adversary (which lies within InstaHide's threat model). First, we assume that the adversary has knowledge of the \emph{distributions} of the private data $X$ and public data $P$. With this knowledge alone, our attack succeeds in recovering the mask $\sigma$, thereby leaking visually-identifiable content of mixed images.
Second, to recover mixed images from a single encoding, we further assume that the adversary has full knowledge of the public dataset $P$.
While this latter assumption is strong, the success of our attack illustrates that if InstaHide is to provide any security even when releasing a \emph{single} encoding, then this security must partially rely on the secrecy of the ``public'' mixing data $P$. 

Our attack proceeds in two steps (with details deferred to below). First, we train a Generative Adversarial Network~\cite{goodfellow2014generative} to learn to ``re-colorize'' \cite{zhang2016colorful} encoded images. That is, we learn the mapping $abs(x) \mapsto x$ where $x$ is a mixture of $k$ images. Learning this mapping requires some prior on the distribution of private data $X$ and mixing images $P$.
Then, we simply compute the \emph{image similarity} of the mixed image with all public images and recover the mixed public images via this simple process (the complexity of this step is linear in $|P|$).

\paragraph{Evaluation.}
We evaluate this attack on CIFAR-10 for an InstaHide scheme with $k=4$. Since a single encoding is released per private image, we mix each private image (from the first $100$ examples in the CIFAR-10 test set), with $3$ images from a public set $P$ containing the remaining $9{,}900$ test samples.
The outputs of our two-stage attack are shown in Figure~\ref{fig:attack_single_encoding}. 

We first train a GAN to learn the mapping $abs(x) \mapsto x$ where $x$ is a mixture of $k=4$ images from the CIFAR-10 training set. Our approach borrows from the use of GANs to colorize grayscale images.\footnote{\url{https://github.com/karoly-hars/GAN_image_colorizing}.} Given the absolute value of a mixed image $abs(x)$, the generator is trained to output a mask $\hat{\sigma} \in [-1, 1]$ so that $abs(x) \circ \hat{\sigma}$ is indistinguishable (to the discriminator) from unmasked mixed images.
In a majority of cases, the GAN re-coloring successfully recovers most of the random mask $\sigma$.

In the second step, given a re-colored mixed image $x$, we  iterate over the public dataset $P$ and compute, for each public image $p$, the \emph{Structural Similarity Index}, $\texttt{SSIM}(x, p)$~\cite{wang2004image}. We select the public image with highest similarity, subtract it from the mixture (we simply ``guess'' that the mixing weight is $\lambda=\frac{1}{k}$), and recurse. That is, we recompute the structural similarity with the remaining public images and repeat until we have subtracted $3$ public images.
This step could potentially be improved by \emph{learning a similarity function}, as we did in Section~\ref{subsec:clustering} for our attack on the InstaHide challenge.

The success of the second step is contingent on the first.
Given an accurate re-colorization, subtracting the public images with highest similarity to the mixture recovers a near-perfect copy of the private image.
For the $100$ encodings we generated, our attack recovers the $3$ public mixing images in $69\%$ of cases, and at least $2/3$ in $85\%$ of cases.

\subsubsection{Attacking InstaHide with a Larger MixUp-$k$}

Recall that the parameter $k$ in InstaHide controls the number of
total images mixed to form one encoded image. The authors argue that larger values
of $k$ result in stronger versions of the scheme.
Specifically the authors claim breaking InstaHide requires
$O(|P|^k)$ work.
Our attack above breaks InstaHide for the setting $k=6$, however as this is a
security parameter it is reasonable to ask if larger values of $k$ would prevent
our proposed attack.

We find it would not. 
Surprisingly, we find that as $k$ gets larger our reconstruction attack becomes \emph{better}.
In Equation~\ref{eqn:min2} we treat the noise $\sigma$, which is only present because of
the public images, as pointwise Gaussian noise.
When $k=6$ this is already an acceptable approximation and the attack succeeds.
But as $k$ grows larger, this approximation gets better and better.
In fact, for $k \to \infty$ we should expect that
the average over all public images will result in no noise.

\section{Conclusion}

Training neural networks while preserving data privacy is of clear
importance in many settings \cite{shokri2017membership, melis2019exploiting}.
In principle, training models with provable privacy guarantees is possible: secure multiparty computation \cite{chase2017private,mohassel2017secureml,wagh2019securenn} or fully homomorphic encryption~\cite{gentry2009fully, hesamifard2017cryptodl} can provide provable cryptographic guarantees on the confidentiality of user data during training, and differential privacy \cite{dwork2014algorithmic,shokri2015privacy,abadi2016deep} can bound the statistical leakage of training data for the final model.

As these provable guarantees can come at a high cost in performance and accuracy, recent work has proposed alternative instance-encoding schemes that aim to offer strong privacy guarantees with little overhead.
Instantiations of these proposals, such as InstaHide \cite{huang2020instahide}, often lack rigorous notions of privacy and rely on ad-hoc security arguments.

We have formalized natural (cryptographic) privacy notions for instance encoding schemes, and have proven strong barriers against achieving these. Specifically, we have shown that any encoding scheme that allows for training accurate models cannot provide similar indistinguishability guarantees as MPC.

We have thus further asked whether existing instance-encoding schemes satisfy weaker privacy notions, in particular a very weak notion of security against reconstruction attacks.
We have shown successful reconstruction attacks on InstaHide~\cite{huang2020instahide}, and in particular we have succeeded in fully breaking the challenge posted by the authors. 
Our attacks directly contradict the heuristic privacy arguments that underlie the InstaHide construction.
As similar constructions underlie other recent proposals for private training~\cite{raynal2020image} and inference~\cite{liudatamix}, these heuristic schemes can likely be defeated by similar attacks.

The goal of privately training neural networks without sacrificing performance is notable, and we hope it will be achievable in the future.
Yet, to enable meaningful progress, proposed schemes should strive to provide precise and \emph{falsifiable} privacy claims, in place of ad-hoc security arguments.

\section*{Acknowledgements}
We thank Shuang Song, the InstaHide Authors, and the anonymous reviewers for feedback on early drafts of this paper.

This paper was supported in part by DARPA under Agreement No. HR00112020026, AFOSR Award FA9550-19-1-0200, NSF CNS Award 1936826, NSF grants CNS-1936799 and CCF-1910681, and research grants by the Sloan Foundation, Visa Inc., and Center for Long-Term Cybersecurity (CLTC, UC Berkeley). Any opinions, findings and conclusions or recommendations expressed in this material are those of the author(s) and do not necessarily reflect the views of the United States Government or DARPA.

\typeout{}
\bibliographystyle{IEEEtran}
\bibliography{IEEEabrv,paper}




\appendix

\section{Proofs} \label{sec:proofs}
\subsection{Proof of Theorem \ref{thm:dataset_encoding}}
\label{apx:proof1}

\begin{proof} We first show that the encodings of datasets sampled from $D_{c_1}$ and $D_{1-c_1}$ are distinguishable with advantage $0.99-2\delta(n)$, by a distinguishing algorithm $q$. The algorithm $q$ gets $\tilde{S}$ which is either the encoding of a dataset sampled from $D_{c_1}$ or $D_{1-c_1}$. Then it trains a model $h$ by applying $L$ on $\tilde{S}$. Then it queries the model $h$ on the test set on 1000 samples from $D_{c_1}$. 
Since the training accuracy of $h$ should be better than $0.51$ with respect to either $c_1$ or $1-c_1$ with probability at least $1-\delta(n)$, the algorithm $q$ can distinguish the two cases by looking at the predictions of the trained model on the examples   In particular, if the predictions were mostly agreeing with $c_1$ the adversary outputs $1$ otherwise it outputs $0$. Specifically, conditioned on the model trained being $0.51$ correct on both datasets sampled from $c_1$ and $c_2$, the algorithm would be able to distinguish correctly with probability at least $0.99$ using a Chernoff bound.  Then applying a union bound we can bound the success of the algorithm by $0.99-2\delta(n)$. 
\begin{equation}\label{eq1}
\big|\Pr[q(E(D^n_{c_1}))=1] - \Pr[q(E(D^n_{1-c_1}))=1]\big| \geq 0.99-2\delta(n)
\end{equation}

So far, we have shown that an algorithm can distinguish between the encoding of $D_{c_1}^n$ and $D_{1-c_1}^n$. But note that we still do not have a real attack as the datasets sampled from these distributions are not labeled according to the same concept function. In the rest of the proof we see how we can use Inequality \ref{eq1} to prove that there are at least two distributions that are labeled according to the same concept function and that their encodings are still distinguishable. 

To prove this, we use three hybrid arguments. We construct two distributions $D_{a}$ and $D_{b}$ as follows. Let $D_{a}$ be a distribution consisting of two parts $D_{a}\equiv \frac{1}{2} D_{a_1} + \frac{1}{2} D_{a_2}$ where 
\begin{align*}
D_{a_1}&=(D,0)\mid c_1(D)=c_2(D)=0\\
D_{a_2}&=(D,1)\mid c_1(D)=c_2(D)=1 \;.
\end{align*}
We also construct $D_{b} =\frac{1}{2}D_{b_1} + \frac{1}{2}D_{b_2}$ such that 
\begin{align*}
D_{b_1} &=(D,0) \mid c_1(D)=1\wedge c_2(D)=0\\
D_{b_2} &= (D,1) \mid c_1(D)=0 \wedge c_2(D)=1 \;.
\end{align*}
%

Note that $D_a$ is constructed in a way that its labels are consistent with both $c_1$ and $c_2$, while $D_b$ is constructed in a way that its labels are consistent with both $c_2$ and $1-c_1$

Now consider an adversary $A_{c_1}$ that wants to distinguish between encodings of datasets sampled from $D_{c_1}$ and $D_{a}$, using the algorithm $q$ described above. We define:
\begin{equation}
\label{eq2}
\Advan(A_{c_1},n) = \big|\Pr[q(E(D^n_a))=1] - \Pr[q(E(D_{c_1}^n))=1]\big|
\end{equation}

Consider an adversary $A_{c_2}$ that tries to distinguish encodings of two distribution $D_b$ and $D_a$ using the algorithm $q$. We define:
\begin{equation}
\label{eq3}
\Advan(A_{c_2}, n) = \big|\Pr[q(E(D^n_a))=1] - \Pr[q(E(D^n_b))=1]\big|
\end{equation}

Similarly, we define $A_{1-c_1}$ and its advantage as follows:
\begin{equation}
\label{eq4}
\Advan(A_{1-c_1}, n) = \big|\Pr[q(E(D^n_b))=1] - \Pr[q(E(D^n_{1-c_1}))=1]\big|
\end{equation}

Putting these together, applying triangle inequality on Equations \eqref{eq2},\eqref{eq3} and \eqref{eq4} we have:
\begin{align*}
&\Advan(A_{c_1}, n) + \Advan(A_{c_2}, n) + \Advan(A_{1-c_1}, n)\\
&\geq \big|\Pr[q(E(D_{c_1}^n))=1] - \Pr[q(E(D_{1-c_1}^n))=1]\big|\\
&\geq 0.99-2\delta. \text{~~~~~(By Inequality \eqref{eq1})}
\end{align*}
Therefore by an averaging argument at least one of the advantages must be at least $\nicefrac{0.99-2\delta(n)}{3}$.

Without loss of generality, assume $\Advan(A_{c_1}) \geq \nicefrac{0.99-2\delta(n)}{3}$.
Now consider a series of $n+1$ distributions $T_0, \dots, T_n$ where $T_1 = D_{c_1}$ and $T_n=D_a$ and for $1\leq i<n$ we have $T_i=\frac{i}{n}\cdot D_a + \frac{(n-i)}{n} D_{c_1}$. Using $n$ hybrid arguments we can show that there exist $i\in n$ such that $q$ would be able to distinguish the encoding of one $T_i$ from $T_{i+1}$. Namely,
$$\big|\Pr[q(E(T_i^n))=1] - \Pr[q(E(T_{i+1}^n))=1]\big|\geq \frac{0.99-2\delta(n)}{3}.$$
Now, we construct the adversary that proves the theorem. Adversary $A$ tries to break $c_1$ and outputs $T_i$ as the distribution of samples. Then, for the two challenge points, the adversary sample $(x_0,y_0)$ and $(x_1,y_1)$ jointly by first selecting a random bit $b$ for the label and setting $y_0=y_1=b$ and then sampling $(x_0,x_1)$ from $(D|c_1(D)=b, D|c_2(D)=b \wedge c_1(D)=b)$.

This way of sampling ensures that the label of the two challenge samples are labeled the same according to $c_1$. 
\end{proof}

\subsection{Proof of Theorem \ref{thm:rich}}
\label{apx:proof2}

\begin{proof}
The adversary first learns a vector of classifiers $G=(h_1,\dots,h_m)$ where each $h_i$ is trained by sampling $n$ examples from $D$ and labeling them according to $c_i$. The adversary would make sure that each $h_i$ has encoded accuracy at least $1-\epsilon(n)$ by repeating the process an expected $\frac{1}{1-\delta(n)}$ number of times. Therefore the expected running time of acquiring such classifiers is $O(m\cdot n/(1-\delta(n))$. Now by linearity of expectation we have
$$\Ex_{\substack{x\gets D\\ \mathbf{x}\gets D^{n-1}\\ \tilde{x}\gets E_X^1(x,\mathbf{x})}} \left[\frac{|F(x)-G(\tilde{x})|}{|F|}\right]\leq \epsilon(n).$$

Therefore, using the Markov inequality, for any $\tau >0$ we have
$$\Pr_{\substack{x\gets D\\ \mathbf{x}\gets D^{n-1}\\ \tilde{x}\gets E_X^1(x,\mathbf{x})}}\left[\frac{|F(x)-G(\tilde{x})|}{|F|}\leq \epsilon(n) + \tau\right] \geq \frac{\tau}{\tau+\epsilon(n)}.$$

Which means if we set $\tau=\gamma-\epsilon(n)$ we get
$$\Pr_{\substack{x\gets D\\ \mathbf{x}\gets D^{n-1}\\ \tilde{x}\gets E_X^1(x,\mathbf{x})}}\left[\frac{|F(x)-G(\tilde{x})|}{|F|}\leq \gamma\right] \geq 1-\frac{\epsilon(n)}{\gamma}.$$

On the other hand, by the $(m,\gamma)$-richness, for any $\tilde{x}$ we have
$$\Pr_{\substack{ x'\gets D}}\left[\frac{|F(x')-G(\tilde{x})|}{|F|}\geq \gamma\right] \geq 0.99.$$

Now for generating the distinguishing samples the adversary $A$ samples two points $(x_0,x_1)$ jointly from $D$ conditioned on both of them having the same label according to $c$. And then when distinguishing, it decides based on $|F(x_0) - G(\tilde{x})|$. If $|F(x_0) - G(\tilde{x})|\geq \gamma$ output $1$ otherwise output $0$. The advantage of this adversary is equal to
\begin{align*}
&\Pr_{\substack{ (x_0,x_1)\gets D^2\\ \mathbf{x}\gets D^{n-1}\\ \tilde{x}\gets E_X^1(x_0,\mathbf{x})}}\left[\frac{|F(x_0)-G(\tilde{x})|}{|F|}\geq \gamma\right]\\~~~~~~&-\Pr_{\substack{(x_0,x_1)\gets D^2\\ \mathbf{x}\gets D^{n-1}\\ \tilde{x}\gets E_X^1(x_1,\mathbf{x})}}\left[\frac{|F(x_0)-G(\tilde{x})|}{|F|}\geq \gamma\right]\geq 0.99 - \frac{\epsilon(n)}{\gamma}.
\end{align*}
This finishes the proof.
\end{proof}

\subsection{Proof of Theorem \ref{thm:single_concept}}
\begin{proof}
In the proof of the theorem, we leverage a learning algorithm $L'$ defined as follows:
\begin{itemize}
    \item{Training:} Given a dataset $S$, train a model $h\gets L(E(S))$.
    \item{Inference:} output a model $h'$ that given an instance $x$, constructs multiple encodings $e_1,\dots, e_k$ using $x$ and then returns the majority vote over all of them $h'(x)=\maj\set{h(e_1),\dots,h(e_k)}$.
\end{itemize}
Having defined this algorithm, we continue designing the attack. The attack algorithm is as follows:
\begin{enumerate}
    \item The adversary first trains a model $h$ using $m$ labeled samples from $D_c$ using the protocol $(E,L)$, and it keeps doing this until the balanced error of the classifier is at most $\epsilon(m)$.
    \item Given a model $h$, construct a classifier $h'$ that given an input $x$, first constructs $k=-20\ln(\tau)/\tau^2$ fresh encodings $e_1,\dots, e_k$ and then returns the majority vote $h'(x)=\maj\set{h(e_1),\dots,h(e_k)}$.
    \item The adversary  jointly samples $(x_0,x_1)\gets (D,D)\mid c(x_0)=c(x_1)$, until it finds a pair $(x_0,x_1)$ such that 
    $$\Pr_{\substack{\mathbf{x}\gets D^{m-1}\\\tilde{x}_0\gets E_X^1(x_0,\mathbf{x})}}[h(\tilde{x}_0) \neq c(x_0)]\geq 1/2 - \tau/2.$$
    and 
    $$\Pr_{\substack{\mathbf{x}\gets D^{m-1}\\ \tilde{x}_1\gets E_X^1(x_1,\mathbf{x})}}[h(\tilde{x}_1)=c(x_1)] \geq 1- \epsilon(m)-\tau/2$$
    \item The adversary outputs $x_0$ and $x_1$, and receives a fresh encoding $u$ of $x_b$ for a random $b$. Then adversary outputs $1$ if $h(u)=c(x_0)$ and $0$ otherwise.  
\end{enumerate}
First lets see what is the advantage of the adversary if it can successfully find the pair $(x_0,x_1)$. The advantage is equal to
\begin{align*}
&\Big|\Pr_{\substack{\mathbf{x}\gets D^{m-1}\\ \tilde{x}_0\gets E_X^1(x_0,\mathbf{x})}}[h(\tilde{x}_0)=c(x_0)] \ \ -\hspace{-8pt}\Pr_{\substack{\mathbf{x}\gets D^{m-1}\\ \tilde{x}_1\gets E_X^1(x_1,\mathbf{x})}}[h(\tilde{x}_1)=c(x_0)]\Big|\\
&\leq \frac{1}{2}-\epsilon(m) - \tau.
\end{align*}
Now we prove that either we have that the error of $h'$ is less than $\tau$ or the adversary can successfully find $(x_0,x_1)$ in polynomial time.
We do this by assuming that $h'$ has error larger than $\tau$ and then proving that adversary can find $(x_0,x_1)$.
Define an event $Z(x)$ for $x\in \cX$ such that $Z(x)=1$ if we have
$$\Pr_{\substack{\mathbf{x}\gets D^{m-1}\\ \tilde{x}\gets E_X^1(x,\mathbf{x})}}[h(\tilde{x})\neq c(x)]\leq 1/2-\tau/2.$$
If for some $x$ we have $Z(x)=1$ 
then using the Chernoff-Hoeffding bound we have $\Pr[h'(x)\neq c(x)] \leq  \tau/4$. Hence, since the error of $h'$ is larger than $\tau$, we have $\Pr_{x\gets D}[Z(x)=0] \geq \tau/2 $. Therefore, there exists a label $y\in\set{0,1}$ such that $\Pr_{x\gets D \mid c(x)=y}[Z(x)=0] \geq \tau/2 $.

Also define an event $W(x)$ for $x\in \cX$ such that $W(x)=1$ if:
$$\Pr_{\substack{\mathbf{x}\gets D^{m-1}\\ \tilde{x}\gets E_X^1(x,\mathbf{x})}}[h(\tilde{x})=c(x)]\leq 1-\epsilon(m)-\tau/2.$$

Since the balanced error of $h$ on encodings is less than $\epsilon(m)$ we have $\Pr_{x\gets D\mid c(D)=y}[W(x)=0]\geq \tau/2.$
Therefore, the probability that $\Pr_{(x_0,x_1)\gets D^2}[Z(x_0)=0 \wedge W(x_1)=0 \wedge c(x_0)=c(x_1)=y]\geq \tau^2/8$. Thus, the adversary can find a pair $(x_0,x_1)$ by sampling $8/\tau^2$ number of samples in expectation.

Putting things together, we have shown that either the advantage or the adversary or the accuracy of $h'$ is high. To finish the proof, we need to calculate the running time of the adversary. The first step of the attack requires  $O(m/\delta(m))$ time. The second step of the attack just requires writing the description of $h'$ which takes constant time. The third step of the attack requires $O(1/\tau^2)$ samples and for each samples we need $O(m)$ time to calculate the events $Z$ and $W$ which makes the running time of the third step $O(m/\tau^2)$ in expectation. Therefore the adversary's running time is $O(m/\tau^2 + m/\delta(m))$ in expectation.

We should also describe the efficiency of the learning algorithm generating $h'$ and also the efficiency of $h'$ itself. Note that although $h'$ is a randomized algorithm as described, we can use standard de-randomization techniques to make it deterministic without losing its accuracy. Then, to run $h'$, one needs to spend $O(1/\tau^3)$ time to calculate the encodings and take the majority. Each encoding takes $O(m)$ time, so overall, the running time of $h'$ with oracle access to $h$ is $O(m/\tau^3)$.

 

\end{proof}

\section{Pixel-Perfect Break InstaHide due to Implementation Flaws}
\label{apx:prng_attack}

The attacks in Section~\ref{sec:instahide_attack} and \ref{sec:instahide_parameters} break the algorithmic foundation of InstaHide, and any implementation
of InstaHide would be vulnerable to these attacks.
We additionally discovered several weaknesses in the \emph{implementation}
of InstaHide that allow us to achieve a \textbf{pixel perfect} reconstruction
of the original dataset.
These implementation weaknesses are \emph{not} fundamental to InstaHide,
and can be easily be corrected; nevertheless, we describe this attack for
completeness. 

As the authors of InstaHide did not release the \emph{ground truth} images for their challenge, this attack also serves as a comparison point for our other (implementation-independent) attacks.
To ensure that this attack does not taint the results of the attacks we developed
in prior sections, we developed this attack only \emph{after} completing all other
aspects of this paper.

At a high level, this attack exploits two weaknesses in the implementation of InstaHide
(and of the InstaHide Challenge):
\begin{itemize}
    \item InstaHide masks each encoded image with a random mask $\sigma$.
    However, instead of using a cryptograhpically secure random number
    generator the implementation calls \texttt{torch.random}, 
    and \texttt{numpy.random}, which
    uses a Mersenne Twister \cite{matsumoto1998mersenne}.
    \item The InstaHide Challenge releases the encoded dataset where each
    pixel is represented as a 32-bit floating point number, $4\times$
    more precision than typical 8-bit integers used to represent images.
\end{itemize}

\subsection{PRNG State Extraction}

Pseudo random number generators (PRNG), work by
maintaining a state vector $v$. When calling the generator, a deterministic function is applied to the current state to yield a new number to output, and an updated state.
Critically, if initialized with the same state, a PRNG will generate the same
output sequence.

The InstaHide implementation uses a Mersenne Twister \cite{matsumoto1998mersenne} PRNG,
the default random number generator in NumPy, in most of its computations.
This includes
the randomness in the encoding, including selecting which original images
will be used to generate each encoded image, generating the $\lambda$ values,
choosing which public images to mix into the private images, and 
generating the random masks $\sigma$. This PRNG is \emph{not} intended for security-sensitive purposes.

We extract the PRNG state via brute force search of the $2^{32}$ possible initial
seeds.\footnote{If this was computationally intractable then stronger
mathematical analysis would allow us to recover the complete state \cite{kelsey1998cryptanalytic}.}
To do this we implement an efficient test that, given a potential PRNG seed, allows
us to determine if the seed was correct.
This allows us to check if any particular seed is correct in roughly $0.1$ milliseconds.
We then repeat this check for each of the $2^{32}$ possible seeds.
This takes $120$ CPU hours, which we parallelize across $100$ cores to obtain the solution in a little over an hour.

Once we extract the PRNG seed, we can use it to compute the exact mapping
$\phi$, the exact values of $\lambda$, and, most impotantly, allows us to
\textbf{undo the encryption operation} of multiplication by $\sigma$.
Note that if InstaHide only released $abs(e)$ for each encoded image $e$,
this attack would not be possible because the information would be truly destroyed.

However, because the authors insist on making an analogy to encryption (and instance
hiding) by multiplying by a random $\{-1,1\}^d$ vector, it is possible to 
``decrypt'' the original images and recover the encoded images without sign information
missing.
This demonstrates that even two mathematically identical techniques can have very different failure modes in practical implementations.

\subsubsection{High-Fidelity Image Reconstruction}

\begin{figure}[b]
    \centering
    \includegraphics[width=.9\columnwidth]{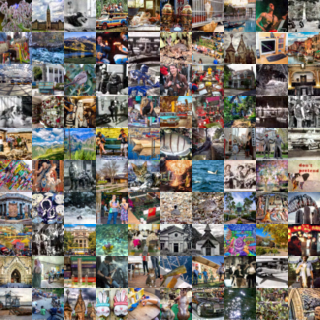}
    \caption{Reconstruction of each of the $100$ images in the InstaHide Challenge,
    using the improved PRNG cryptanalytic attack that exploits implementation
    weaknesses in InstaHide.}
    \label{fig:improved}
\end{figure}

Given all of this information ($\phi$, $\lambda$, and $E$ without sign flipping),
the reconstruction attack from Section~\ref{sec:partialthing} applies directly.
Figure~\ref{fig:instahide_break} shows the result of this attack on the InstaHide challenge compared
to the images we extract using the prior attack.
All $100$ reconstructed images are given in Figure~\ref{fig:improved}.

\subsubsection{Pixel-Perfect Refinement}

We are able to make one final improvement that allows us to recover a
\emph{pixel perfect} reconstruction when given access to the public dataset.
Because we have reverse engineered the PRNG seed, it turns out that not only do
we get access to the function $\phi$ but we can even determine \emph{which}
public images were used in each encoded image---because these values are
determined using the same PRNG. As a result of this, we now have an \emph{over-determined} system of equations.
By replacing the noise value $\sigma$ from Equation~\ref{eqn:min2} with the actual
public images, this reduces the number of free variables to just $M \cdot d$ 
when there are $M$ original images of dimension $d$.
Because the number of encoded images is greater than the number of 
original images (and in practice $50\times$ as many for the Challenge)
we can perfectly solve for the reconstruction.

Unfortunately we are unable to mount this attack on the actual InstaHide
Challenge: the authors do not release the public dataset of
the challenge dataset.
However, we have confirmed this attack on CIFAR-10 and it works
as expected.
\section{Additional Figures}
\label{apx:figures}

\begin{figure}[!h]
    \centering
    \includegraphics[width=.9\columnwidth]{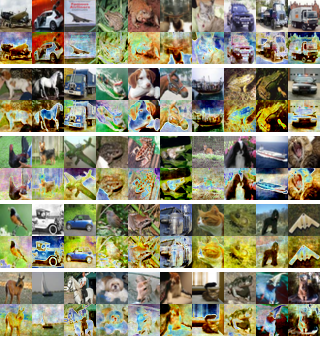}
    \caption{Reconstruction of the first $50$ images in the CIFAR-10 encoded dataset.
    In each pair of rows, the upper image is the original and the lower image
    is the reconstruction.}
    \label{fig:cifar10_all}
\end{figure}

\begin{figure}[!h]
    \centering
    \includegraphics[width=.9\columnwidth]{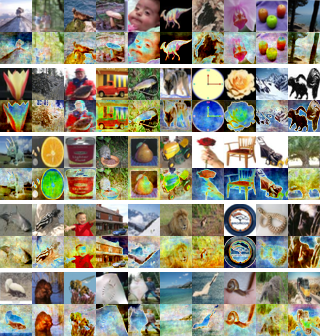}
    \caption{Reconstruction of the first $50$ images in the CIFAR-100 encoded dataset.
    In each pair of rows, the upper image is the original and the lower image
    is the reconstruction.}
    \label{fig:cifar100_all}
\end{figure}

\begin{figure}[!h]
    \centering
    \includegraphics[width=.9\columnwidth]{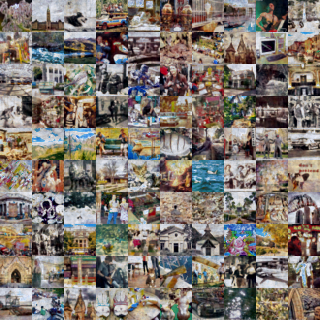}
    \caption{Reconstruction of each of the $100$ images in the InstaHide Challenge
    using the fully general attack.}
    \label{fig:all}
\end{figure}

\end{document}